\documentclass[preprint]{aastex}
\usepackage{hyperref}
\usepackage{graphicx}
\usepackage{epstopdf}
\usepackage{lineno}
\usepackage{lscape}
\newcommand{\CGRO}{\textit{CGRO}}
\newcommand{\Fermi}{\textit{Fermi}}
\newcommand{\Swift}{\textit{Swift}}
\newcommand{\INTEGRAL}{\textit{INTEGRAL}}
\newcommand{\Suzaku}{\textit{Suzaku}}
\newcommand{\AGILE}{\textit{AGILE}}
\newcommand{\MAXI}{\textit{MAXI}}
\newcommand{\rb}[1]{\raisebox{1.5ex}[-1.5ex]{#1}}

\shorttitle{\Fermi\ 3$^{\rm rd}$ GBM Gamma-Ray Burst Catalog}
\shortauthors{P.\,N.\,Bhat et al.}

\begin{document}

\title{The 3$^{\rm rd}$ \Fermi\ GBM Gamma-Ray Burst Catalog: \\
  The First Six Years}

\author{
P.~Narayana~Bhat\altaffilmark{1,12},
Charles A.~Meegan\altaffilmark{1},
Andreas~von~Kienlin\altaffilmark{3},
William S.~Paciesas\altaffilmark{2},
Michael~S.~Briggs\altaffilmark{1,12},
J.~Michael~Burgess\altaffilmark{16,17},
Eric~Burns\altaffilmark{1},
Vandiver~Chaplin\altaffilmark{1,11},
William~H.~Cleveland\altaffilmark{2},
Andrew~C.~Collazzi\altaffilmark{8},
Valerie~Connaughton\altaffilmark{2,12},
Anne~M.~Diekmann\altaffilmark{9},
Gerard~Fitzpatrick\altaffilmark{1,6},
Melissa~H.~Gibby\altaffilmark{9},
Misty~M.~Giles\altaffilmark{9},
Adam~M.~Goldstein\altaffilmark{7},
Jochen~Greiner\altaffilmark{3,10},
Peter~A.~Jenke\altaffilmark{1,12},
R.~Marc~Kippen\altaffilmark{14},
Chryssa~Kouveliotou\altaffilmark{4},
Bagrat~Mailyan\altaffilmark{1},
Sheila~McBreen\altaffilmark{6},
Veronique~Pelassa\altaffilmark{1,13},
Robert~D.~Preece\altaffilmark{12},
Oliver~J.~Roberts\altaffilmark{6},
Linda~S.~Sparke\altaffilmark{5},
Matthew Stanbro\altaffilmark{1},
P\'{e}ter~Veres\altaffilmark{1},
Colleen~A.~Wilson-Hodge\altaffilmark{7},
Shaolin~Xiong\altaffilmark{15},
George~Younes\altaffilmark{4},
Hoi-Fung~Yu\altaffilmark{3,10}
and Binbin Zhang\altaffilmark{1}
}
\altaffiltext{1}{The Center for Space Plasma and Aeronomic Research (CSPAR), University of Alabama in Huntsville, 320 Sparkman Drive, Huntsville, AL 35805, USA}
\altaffiltext{2}{Universities Space Research Association, 320 Sparkman Drive, Huntsville, AL 35805, USA}
\altaffiltext{3}{Max-Planck-Institut f\"{u}r extraterrestrische Physik, Giessenbachstrasse 1, 85748 Garching, Germany}
\altaffiltext{4}{Department of Physics, The George Washington University, 725 21st Street NW, Washington, DC 20052, USA}
\altaffiltext{5}{Astrophysics, Science Mission Directorate, NASA HQ, 300 E Street SW, Washington, DC 20546, USA}
\altaffiltext{6}{School of Physics, University College Dublin, Belfield, Stillorgan Road, Dublin 4, Ireland}
\altaffiltext{7}{ZP12 Astrophysics Office, NASA-Marshall Space Flight Center, Huntsville, AL 35812, USA}
\altaffiltext{8}{SciTec Inc., 100 Wall St., Princeton NJ, 08540, USA}
\altaffiltext{9}{Jacobs Technology, Inc., Huntsville, Alabama, USA}
\altaffiltext{10}{Excellence Cluster Universe, Technische Universit\"at M\"unchen, Boltzmannstr. 2, 85748, Garching, Germany}
\altaffiltext{11}{Vanderbilt University Institute of Imaging Science, 1161 21$^{st}$ Avenue South, Medical Center North, AA-1105, Nashville, TN 37232, USA}
\altaffiltext{12}{Department of Space Science, University of Alabama in Huntsville, 320 Sparkman Dr., Huntsville, AL 35805}
\altaffiltext{13}{Fred Lawrence Whipple Observatory, 670 Mount Hopkins Road, Amado AZ 85645, USA}
\altaffiltext{14}{Los Alamos National Laboratory, MS B244, P.O. Box 1663, Los Alamos, NM 87545, USA}
\altaffiltext{15}{Key Laboratory of Particle Astrophysics, Institute of High Energy Physics,19B Yuquan Road, Beijing, 100049, China}
\altaffiltext{16}{The Oskar Klein Centre for Cosmoparticle Physics,
AlbaNova, SE-106 91 Stockholm, Sweden}
\altaffiltext{17}{Department of Physics, KTH Royal Institute of Technology, AlbaNova University Center, SE-106 91 Stockholm, Sweden}

\begin{abstract}

Since its launch in 2008, the \Fermi\ Gamma-ray Burst Monitor (GBM) has triggered and located on average approximately two  $\gamma$-ray bursts (GRB) every three days. Here we present  the third of a series of catalogs of GRBs detected by GBM, extending the second catalog by two more years, through the middle of July 2014. The resulting list includes 1405 triggers identified as GRBs.  
The intention of the GBM GRB catalog is to provide information to the community on the most important observables of the GBM detected GRBs.  For each GRB the location and main characteristics of the prompt emission, the duration, peak flux and fluence are derived. The latter two quantities are calculated for the 50  -- 300~keV energy band, where the maximum energy release of GRBs in the instrument reference system is observed, and also for a broader energy band from 10 -- 1000~keV, exploiting the full  energy range of GBM's low-energy NaI(Tl) detectors. 
Using statistical methods to assess clustering, we find that the hardness and duration of GRBs are better fitted by a two-component model with short-hard and long-soft bursts, than by a model with three components.
Furthermore, information is provided on the settings and modifications of the triggering criteria and exceptional operational conditions during years five and six in the mission. This third catalog is an official product of the \Fermi\ GBM science team, and the data files containing the complete results are available from the High-Energy Astrophysics Science Archive Research Center (HEASARC).

\end{abstract}


\section{INTRODUCTION}

Since the first  $\gamma$-ray burst (GRB) was observed by the Vela satellite \citep{1973ApJ...182L..85K}, the number of flashes of high-energy radiation that have been detected has increased dramatically, especially since the 1991 launch of the Compton Gamma Ray Observatory and its $\gamma$-ray burst instrument, the Burst and Transient Source Experiment (BATSE) \citep{1992Nature...355..143}. The GBM technique of detecting and locating GRBs is largely based on BATSE \citep{1989GROScienceWS}, which operated from 1991 to 2000. Both instruments employ multiple sodium iodide [NaI(Tl)] detectors to achieve full sky field of view, have on-board burst triggering capability and use the relative count rates to obtain approximate directions to bursts. GBM also includes two bismuth germanate (BGO) detectors that are better suited for the detection of higher energy $\gamma$-ray photons. BATSE, with significantly larger ($20^{\prime\prime}$ in diameter $1^{\prime\prime}$ thick) NaI(Tl) detectors, had better sensitivity, while GBM has a broader energy range and higher data rate.

\Fermi\  was launched on  June 11, 2008 and is now operating successfully in space for more than seven years. GBM's main task is to augment the mission's capability to detect and  locate GRBs as well as to provide broad-band spectral information.  GBM extends the energy range of the main instrument, the Large Area Telescope (LAT: 30~MeV -- 300~GeV) down to the soft  $\gamma$-ray and X-ray energy range (8~keV -- 40~MeV). This allows for observations over more than seven decades in energy.

In the six years of operation since triggering was enabled on 2008 July 12, GBM has triggered 3350 times on a variety of  transient events:  1405 of these are classified as GRBs (in two cases the same GRB triggered GBM twice), 198 as bursts from soft  $\gamma$ repeaters (SGRs), 469 as terrestrial  $\gamma$-ray flashes (TGFs), 795 as solar flares (SFs), 304 as charged particle (CPs) events, and 179 as other events (Galactic sources, accidental statistical fluctuations, or too weak to classify). Table \ref{trigstat1st2ndcat} shows a breakdown of the observed event numbers sorted by the time periods covered by the first GBM burst catalog (2008 July 12 to 2010 July 11) \citep{Pacie12}, the second GBM catalog (2010 July 12 to 2012 July 11) \citep{kie14} and the additional two years included in the current catalog  (2012 July 12 to 2014 July 11) again separated according to the event type. In addition, the number of Autonomous Repoint Requests (ARRs, described in Section \ref{SecTriggDiss} below) and GRBs detected by LAT, observed with high confidence above 100 MeV (and 20 MeV), are given \citep{2013ApJS..209...11A}. The preliminary results of the GBM team analyses for bright bursts and those bursts simultaneously detected by other satellite instruments 
are reported in the Gamma-ray Coordinates Network (GCN) circulars\footnote{ \url{http://gcn.gsfc.nasa.gov/}}, a very effective way of informing the GRB research community of the initial properties of GRBs. Here we present the final results after a careful analysis of the full set of burst data using detector response functions for the best available burst location. This catalog lists for each GRB the location and the main characteristics of the prompt emission, the duration, peak flux and fluence. In addition, the distributions of these derived quantities for the entire 6 year period are also presented. 

The upcoming GBM spectral catalog \footnote{GBM 8 Year Spectroscopy Catalog, in prep.} provides information on the systematic spectral analysis of nearly all GRBs listed in the current catalog, including time-integrated fluence and peak flux spectra.   A new catalog of time resolved spectral analysis of bright GRBs of the first 4 years has also been compiled \citep{yud15}.

In \S\ref{SecDets} and \ref{SecTriggDiss} we briefly describe the GBM detectors and the GBM GRB localization technique together with a description of the onboard triggering algorithms and the path of trigger information dissemination. Furthermore a brief description of the GBM data products is presented. Section \ref{trigger_statistics} reports the GRB trigger statistics of the first six years, comparing them with the triggers on other event classes. Major changes in operational conditions occurring during years 5 $\&$ 6 are also mentioned. A summary of the major steps of the catalog analysis is given in \S\ref{CatAna}. The catalog results are presented in \S\ref{CatRes} and are discussed in \S\ref{Disc}. Finally, in \S\ref{CatSum} we conclude with a summary.

\section{FERMI GAMMA-RAY BURST MONITOR}
\label{gbm}

\subsection{GBM Detectors}
\label{SecDets}

The capability of GBM to detect and locate GRBs in the energy range of the maximum energy release in the observer reference system and to provide energy overlap with  that of the main instrument (LAT) is achieved by the use of two different types of inorganic scintillation detectors. In the energy range from 8~keV to 1~MeV, 12 thallium-doped sodium iodide detectors [NaI(Tl)], each of which is attached to a $5^{\prime\prime}$ photomultiplier tube (PMT), are used. The NaI(Tl) detectors are each 1.27 cm thick by 12.7 cm diameter and are deployed around the spacecraft in such a way that each detector observes the sky at a different inclination, providing visibility of the entire sky unocculted by the Earth. The relative count rates of the NaI detectors,  which have a quasi-cosine response, are used to determine the locations of triggered GRBs. The location of a GRB is calculated by comparing the measured background-subtracted count rates in individual detectors with a lookup table, containing a list of relative detector rates for a grid of simulated sky locations of  $\gamma$-ray point sources.

The on-board and on-ground lookup tables have resolutions of 5 degrees and 1 degree, respectively. With this method the limiting accuracy is approximately 8 degrees for on-board locations and approximately 4 degrees for on-ground locations. A detailed investigation of the GBM location accuracy can be found in \citet{Con15}. 

For the detection of the prompt  $\gamma$-ray emission in the MeV-range, between $\sim 200$~keV and 40 MeV, detectors employing the high density scintillation crystal Bismuth Germanate (BGO) are used. Two detectors using large cylindrical BGO crystals, 12.7 cm diameter by 12.7 cm thick, each coupled to two $5^{\prime\prime}$ photomultipliers, one on each end, are mounted on opposite sides of the spacecraft, allowing observations of the full unocculted sky and providing spectral information up to tens of MeV regime for all GBM detected bright and hard GRBs. The GBM instrument is described in more detail in \citet{2009ApJ...702..791M}.

\subsection{Triggering and Post-trigger Operations}
\label{SecTriggDiss}

The GBM Flight Software (FSW) continuously monitors the detector count rates to detect GRBs and other short-timescale transients.
A  burst  trigger  occurs  when  the  FSW  detects  an increase in the count rates of two or more NaI(Tl) detectors above a preset but adjustable threshold specified in units of the standard deviation of  the  background  rate.  The  background  rate  is  an  average rate accumulated over the previous 17 seconds, excluding the most recent 4 s. Energy ranges are confined to combinations of the eight channels of the CTIME  data \citep{2009ApJ...702..791M}. Trigger timescales may be defined as any multiple of 16 ms up to 8.192 s. Except for the 16 ms timescale, all triggers include two phases offset by half of the accumulation time, which has been suggested to be optimal \citep{2002ApJ...578..806B}. A total of 120 different triggers can be specified, each with a distinct threshold. The  trigger  algorithms  currently  implemented  include  four
energy  ranges:  the  BATSE  standard  50 - 300  keV  range,  25 - 50 keV to increase sensitivity for SGRs and GRBs with soft spectra, $>$ 100  keV,  and $>$ 300  keV  to increase  sensitivity  for hard GRBs and TGFs \citep{2011JGRA..116.7304F}. Ten timescales, from 16 ms to 8.192 s in steps of a factor 2, are implemented in the 50 - 300 keV range and the 25 - 50 keV range. The $>$ 100 keV trigger excludes the 8.192 s timescale, and the $> $300 keV trigger has only four timescales, viz. 16, 32, 64 and 128 ms. The concept of the trigger algorithm was adopted from BATSE, but with added algorithms running in parallel. The large number of algorithms and flexibility has made it is possible to investigate whether the population of GRBs observed by BATSE was actually biased by the latter's limited number of trigger algorithms. This also adds an improvement in the GBM trigger sensitivity \citep{2002ApJ...578..806B, 2004AIPC..727..688B}. The standard setting of the offset is half the timescale of the original algorithm as mentioned above.  A summary of the actual settings (by July 2014) and the changes in the first six years of the mission is shown in Table \ref{trigger:criteria:history}.

Since GBM triggers on a variety of astrophysical transients in addition to GRBs, the FSW performs an automatic event classification by using a Bayesian approach that takes into account  (i) the event localization, (ii) spectral hardness, and (iii) the spacecraft geomagnetic latitude \citep{2009ApJ...702..791M}.  This information is very important and useful for the automated follow up observations. Furthermore the capabilities of the instrument to detect events other than GRBs were improved by tuning dedicated trigger algorithms using refined Bayesian priors.

Following a trigger some important parameters that are needed for rapid ground-based follow-up observations, {\em i.e.}, onboard preliminary localization, event classification, burst intensity and background rates, are downlinked as TRIGDAT data by opening a real time communication channel through the Tracking and Data Relay Satellite System (TDRSS). In addition, these data are used in near real-time by the Burst Alert Processor (BAP), redundant copies of which are running at  Goddard Space Flight Center (GSFC) and at the GBM Instrument Operations Center (GIOC) at the National Space Science and Technology Center (NSSTC) in Huntsville, Alabama. Relative to the GBM FSW, the BAP provides improved locations, since it uses a finer angular grid (1 degree resolution) and accounts for differences in the burst spectra and more accurately for atmospheric and spacecraft scattering. Users worldwide are informed within seconds about the flight and automatic ground locations and other important parameters by the automatic dissemination of notices via the GCN as mentioned before.
The GBM burst advocates (BA), working in alternating 12 hr shifts at the GIOC and collaborating institutions in Europe, including the operations center MGIOC at the Max Planck Institute for Extraterrestrial Physics (MPE) in Garching, Germany, use the TRIGDAT data to promptly confirm the event classification and generate refined localizations by applying improved background models and detector response functions.  Unless a more precise localization of the same GRB has been reported by another mission, a GCN notice with the final position and classification is sent out by the BA. In addition, the BAs compute preliminary durations, peak fluxes, fluences and spectral parameters, and report the results in a GCN circular in case of a bright event or a GRB that was already detected by another instrument.

The GBM FSW promptly notifies the LAT of trigger times and locations of triggered GRBs as well as their preliminary classifications in order to launch dedicated onboard burst search algorithms for the detection of possible high-energy emission. In case of a sufficiently intense GRB, which exceeds a preset threshold for peak flux or fluence, a request for an ARR of the spacecraft is transmitted to the LAT and forwarded to the spacecraft FSW. Acceptance of the ARR by the spacecraft FSW initiates a special observation mode that maintains the burst location in the LAT field of view for an extended duration (currently 2.5 hours, subject to Earth limb constraints), to search for delayed high-energy emission. Table \ref{trigstat1st2ndcat}  lists the total number of ARRs which occurred in the first six years and the table~\ref{main_table} identifies those GRBs for which the GBM FSW issued an ARR.

The continuous background count rates recorded by each detector are downlinked as two complementary data types, the 256 ms high temporal resolution CTIME data with 8 energy channels and the low temporal resolution (4 s) CSPEC data with full spectral resolution of 128 energy channels that are used for spectroscopy.
The lookup tables (LUTs) are used by the GBM FSW to define the boundaries of the CTIME and CSPEC  spectral energy channels from the 4096 ADC channels. There are two CSPEC LUTs, one for the NaI(Tl) detectors  and  one  for  the  BGO  detectors. The current LUTs are pseudo-logarithmic so that spectral channel widths are
commensurate  with  the  detector energy resolution  as  a  function  of  $\gamma$-ray energy. In case of an on-board trigger, the temporal resolutions of CTIME and CSPEC data are increased to 64~ms and 1~s, respectively, a mode lasting nominally for 600 s after the trigger time.

Moreover, high temporal and spectral resolution data are downlinked for each triggered event. These time-tagged event (TTE) data consist of individual photon arrival times with 2~$\mu$s temporal resolution and 128 channel spectral resolution from each of the 14 GBM detectors, are recorded for 300~s after and about 30~s before each trigger. The benefit of this data type is the flexibility to adjust the temporal resolution to an optimal value with sufficient statistics during analysis. Beginning on November 27,  2012, TTE data have been generated continuously throughout the orbit except during \Fermi\  passage through South Atlantic Anomaly (SAA, see section \ref{fswupgrade} for more details).

\subsection{Trigger Statistics}
\label{trigger_statistics}

\begin{figure}[h]
\begin{center}
\includegraphics[scale=0.80, angle=-90,totalheight=0.45\textheight]{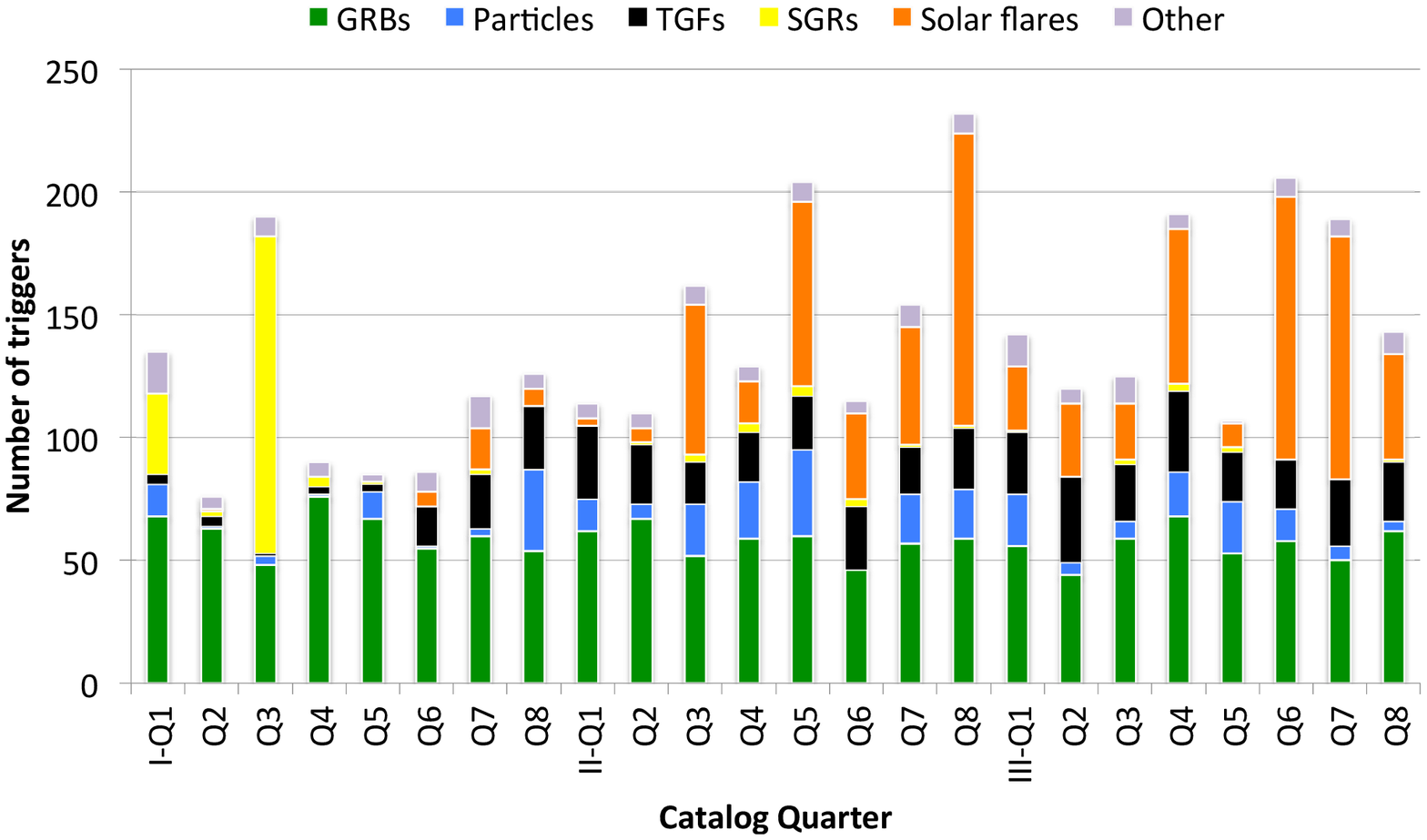}
\caption{\label{quarterlytrigstat} The quarterly trigger statistics over the first six years of the GBM mission. The total number of triggers in the time period from  July 12, 2008 to  July 11, 2014 are shown.  The first quarter starts at July 12, 2008 (I-01) and ends at October 11, 2008 and so on until July 11, 2014. The different types of events triggering GBM are classified as shown at the top.  The 'Other' category includes accidental triggers and those that are too weak to locate or unambiguous in nature, as well as transient burst activity from galactic sources such as Cyg X-1 and V404 Cyg.}
\end{center}
\end{figure}

The GBM instrument, which is primarily designed to detect cosmic GRBs, additionally detects bursts originating from other cosmic sources, such as Solar Flares (SFs) and SGRs, as well as extremely short but spectrally hard TGFs observed from the Earth's atmosphere, which have been associated with lightning events in thunderstorms. Table \ref{trigstat1st2ndcat}  summarizes the numbers of triggers assigned to these additional event classes, showing that their total number is of the same order as the total number of triggered GRBs. Approximately 10\% of the triggers are due mostly to cosmic rays or trapped particles; the latter typically occur in the entry or exit regions of the SAA or at high geomagnetic latitude. In rare cases, outbursts from known Galactic sources have caused triggers. Finally, $\sim 6$\% of the GBM triggers are generated accidentally by statistical fluctuations or are too weak to be confidently classified.
The quarterly trigger statistics over the six years of the mission are graphically represented in Figure \ref{quarterlytrigstat}. The rate of GRBs is slightly lower in the second two years because at the beginning of 2011 July triggers were disabled during times when the spacecraft was at high geomagnetic latitude. The McIlwainL parameter, which is used by the FSW as a threshold for disabling triggering, has been raised from 1.58 to 1.9 since year 5 while the McIlwain L threshold for disabling ARRâ has been retained at 1.58. 

It is evident from Figure~\ref{quarterlytrigstat} that the major bursting activity from SGR sources took place in the beginning of the mission, mainly in 2008 and 2009. In addition to emission from previously known SGR sources \citep{2012ApJ...755..150V,2012ApJ...749..122V,2011ApJ...739...87L}, GBM also detected a new SGR source \citep{2010ApJ...711L...1V}.
It is also obvious from the figure that the rate of monthly detected triggers on TGF events has increased by a factor of $\sim 8$ to about two per week, after the upload of the new  FSW version (V2.6) on November 10, 2009 \citep{2013JGRA..118.3805B}. This version includes additional trigger algorithms that monitor the BGO detector count rates in the 2 -- 40~Mev energy range (see Table \ref{trigger:criteria:history}). This has two advantages: firstly the TGF bursts show very hard spectra up to 40~MeV and hence BGO detectors have a better sensitivity for TGFs, and secondly the deadtime in the NaI(Tl) detectors is much larger for $\gamma$-rays of energy $\ge$1\,MeV\footnote{$\gamma$-rays of energy $>$ 1 MeV will be registered in the overflow channel of the NaI(Tl) spectrum which have 10 times larger deadtime.} \citep{2013JGRA..118.3805B}.

Table \ref{trigstatalgor} summarizes first trigger algorithm that triggered on bursts or flares from the different object classes. Once a trigger has occurred the FSW continues to check the other trigger algorithms and ultimately sends back the information in TRIGDAT data as a list of trigger times for all algorithms that were satisfied. This detailed information was used in \cite{Pacie12} to investigate the apparent improvement in trigger sensitivity relative to BATSE.
It was found that mainly GBM's additional longer trigger timescales triggers ($> 1.024$~s) in the 50 to 300 keV energy range were able to detect GRB events which wouldn't have triggered the BATSE experiment. These observations are confirmed by analyzing the current full 6 year dataset.

Furthermore we ascribe the improved trigger sensitivity of GBM to lower trigger threshold of $4.5\sigma - 5.0\sigma$ (see Table \ref{trigger:criteria:history}) compared to the BATSE settings  of  $5.5\sigma$ \citep[see Table 1 in][]{1999ApJS..122..465P}.
The longest timescale trigger algorithms in the 50 - 300~keV energy range, running at $\sim 16$~s (20, 21) and $\sim 8$~s (18, 19) were disabled in the beginning of the mission (see Table \ref{trigger:criteria:history}), since no  event triggered algorithms 20 \& 21 and only three GRBs triggered on algorithms 18 \& 19.
The algorithms running on energy channel 2 (25 -50 keV) with timescales higher than 128~ms were disabled, since they were mostly triggered by non GRB (and non SGR) events.
The short timescale algorithms in the 25 - 50~keV energy range (22 - 26) were retained, mainly for the detection of SGR bursts, which are short and have soft energy spectra. Even these triggers were disabled during periods of high Solar activity (see table \ref{trigger_modification_history}).
The new algorithms above 100~keV  didn't increase the GRB detection rate. They were disabled with the exception of the shortest timescale algorithms running at 16~ms, particularly suitable for the detection of  TGFs. Table \ref{trigger:criteria:history} clearly shows the capabilities of the newly introduced  ``BGO"-trigger algorithms 116 -119 for TGF detection.

\subsection{GBM Flight Software Upgrade:  Year five and six}
\label{fswupgrade}

As in previous years,  the GBM instrument trigger configuration was temporarily changed in three ways that affected the GRB data: 1) some or all of the trigger algorithms were disabled, and 2) the low-level energy thresholds (LLT) were raised on the Sun-facing detectors (NaI 0-5) and 3) the soft triggers ({\em i.e.} trigger algorithms 22-26) were disabled every weekend Friday 15-20h UT to 13-20h UT Monday for durations anywhere between 60 to 120 hrs. This was to mitigate the generation of excessive TTE data during possible solar activity over the weekends\footnote{A table summarizing the intervals of non-nominal trigger settings is posted at:\\ \url{http://fermi.gsfc.nasa.gov/ssc/data/access/gbm/llt\_settings.html}}.
During years 5-6 \Fermi\ conducted a series of nadir-pointing observations to allow the LAT to detect possible $>$ 100 MeV photons during a TGF. In order to avoid spacecraft reorientation due to a possible ARR from GBM FSW, GRB triggers were disabled during these nadir observations. During these intervals TTE data generation was turned on continuously so that a sensitive search for GBM TGFs coincident with the LAT as well as untriggered GRBs, if any, could be performed. Again, in order to mitigate unacceptably high rates of TTE, the LLTs in the Sun-facing NaI(Tl) detectors were raised above the nominal as summarized in table \ref{trigger_modification_history}.

The hardware design limits the data available to the flight computer for triggering to data binned at temporal resolutions of 16 ms and longer. Hence the onboard triggering of GBM has reduced sensitivity compared to the capabilities of the detectors because the 16 ms minimum resolution of the data used for triggering is much longer than the duration of most TGFs of about 0.1 ms, which adds unnecessary background data and reduces the trigger sensitivity. As mentioned above, the GBM TTE data type records the energies and arrival times of individual photons with 2$\mu$s temporal resolution, 2-5 $\mu$s absolute accuracy, and an energy resolution of 128 pseudo-logarithmic channels. The continuous TTE coverage could enhance the capability of the ground software to detect short untriggered GRBs that could potentially be coincident with a gravitational wave signal detected by LIGO \citep{2015CQGra..32r5003M}. Aforesaid limitations could be circumvented by downlinking the GBM photon data as continuous TTE data and by conducting a ground-based search for TGFs and untriggered GRBs. However, early in the mission producing this data type continuously would have exceeded the original telemetry allocation of GBM. Because of its potential, increased telemetry was provided to support TTE production for a portion of the \Fermi\ orbit each day (FSW V2.6) beginning 9 July 2010.  To use this resource most effectively, continuous (i.e., non-trigger) TTE data were produced in regions where high TGF activity was expected. Polygonal geographic regions were defined. TTE data production was commanded on when \Fermi\  entered one of these regions and commanded off when \Fermi\ exited. This mode of operation continued until 26 November 2012, at which time
the continuous TTE data mode was extended to the entire orbit outside of the SAA. The increased sample size and improved exposure uniformity increased the usefulness of \Fermi\ for studies of TGF properties and TGF-meteorological correlations. The rate of detection of TGFs increased to nearly 850 per year \citep{2013JGRA..118.3805B}. The continuous TTE coverage enhanced the astrophysics capabilities of GBM as well \citep{ 2014PhRvD..89l2004A}.

This new provision of the continuous TTE data production has the potential of producing excessive data that is not very useful during bright Solar Flares. To guard against this, a provision to suspend the TTE data production from the Sun-facing GBM detectors (NaI 0-5 and BGO0) was included in GBM FSW version 2.7, which was uploaded on 25 November 2012. In this version the total count rates from all 14 GBM detectors are continuously monitored by the FSW. If the total rate exceeds a threshold value, the TTE data production from the Sun-facing detectors is suspended while the other detectors continue TTE data production. The TTE data production by the Sun-facing GBM detectors resumes once the total rate falls below the threshold value and remains so for a preset time duration. The FSW parameters that are used to regulate the TTE data production may be changed by FSW commands.

\subsection{Trigger Status Modifications during Year 5 \& 6}

Normal triggering criteria were changed under certain circumstances, such as increased solar activity when a large volume of TTE data are produced.  In order to mitigate the possibility of filling the onboard hard disk, we disabled low energy triggers (trigger algorithms 22-26). However, if the flare is soft and yet very intense it could also trigger on higher energy algorithms during intense Solar Flares. To mitigate such cases we also raised the threshold of the Sun-facing detectors (NaI 0-5)  (see table \ref{trigger_modification_history} for time duration of such changes).

As noted above, during this period \Fermi\ was periodically run in a special mode in which the spacecraft was inverted to view the nadir. 
Each nadir observation lasted for about 2-3 hours (see table \ref{trigger_modification_history}). In order to prevent an unexpected ARR from reorienting the spacecraft during nadir observations, we disabled GRB triggers.  All the times and types of such GBM GRB trigger criteria modifications, are listed in table \ref{trigger_modification_history}.

\section{ANALYSIS RELATING TO THE PRESENT GRB CATALOG}
\label{CatAna}
The GBM GRB catalog analysis process is described in detail in the first and second catalog papers (see appendix of \cite{Pacie12, kie14}). Here we briefly summarize the major analysis steps.

\subsection{Burst Localization and Detector Response Matrix}

The GRB locations listed in Table \ref{main_table} are adopted from the BA analysis results, uploaded to the GBM trigger catalog at the GIOC (with a copy at the \Fermi\ Science Support Center, FSSC\footnote{\url{http://heasarc.gsfc.nasa.gov/FTP/fermi/data/gbm/bursts/}}). Better locations, if any,  for bursts that were detected simultaneously by any other  $\gamma$-ray instrument, such as \Swift  \citep{2005SSRv..120..143B} or INTEGRAL \citep{2003NIMPA.513..118B}, 
or were localized more precisely by the Inter Planetary Network \citep[IPN,][]{2013ApJS..207...39H} are also listed in the table. The determination of a reliable location is quite important since all analysis results  depend on the response files generated for the particular GRB location.

GRB locations shown in Table \ref{main_table} were produced using version 4.15 of the localization code.  The GBM location uncertainties shown in the table are the circular area equivalent of the statistical uncertainty (68\% confidence level).  The source localization algorithm has not changed since the production of the last GRB catalog, and is detailed therein.  The accuracy of the locations has been assessed using a reference sample of 200 GRBs localized by other instruments and 100 GRBs localized through the IPN.  We find the distribution of systematic uncertainties for GRBs is well represented (at 68\% confidence level) by a Gaussian (of standard deviation 3.7\degr) with a non-Gaussian tail that contains about 10\% of GBM-detected GRBs and extends to approximately 14\degr \citep{Con15}. A more complex model suggests that there is a dependence of the systematic uncertainty on the position of the GRB in spacecraft coordinates, with GRBs in the quadrants on the Y-axis better localized than those on the X-axis.  A convolution of the statistical uncertainty with our best current model for the systematic errors produces probability maps reflecting the total uncertainty on a GBM GRB location.   The maps reflect the occasional non-circular shapes of the statistical uncertainty region as well as its area.   They have been packaged into new data products (ASCII, FITS, png) that have been routinely delivered to the FSSC since January 2014 and have now been processed and delivered for the GRBs prior to 2014.  Because the generation of these new data products requires re-running the localization code, which has recently been automated, the actual GRB locations in the GRB catalog at the FSSC may be different from those published in the first two catalogs.  The new locations are consistent, within errors, with the old locations, but the manual processing may use different background and source intervals than those selected in the automatic process.

Detector Response Matrices (DRMs) generated using the General Response Simulation System (GRESS) \citep{2005NCimC..28..797H}, agree  with the data from the detector-level calibrations as a function of angle and energy within better than $\pm $ 5\% for the BGO detectors, and $\pm $ 10\% for the NaI(Tl) detectors. The DRMs, which contain the multivariate effective detection area, include the effects of angular dependence of the detector efficiency, partial energy deposition in the detector, energy dispersion and nonlinearity of the detector, and atmospheric and spacecraft scattering (and shadowing) of photons into the detector, partial or complete blockage of GBM detectors by the LAT or the radiators. They are therefore functions of photon energy, measured (deposited) energy, the direction to the source with respect to the spacecraft, and the orientation of the spacecraft with respect to the Earth. Individual DRMs needed for analysis of the science data were generated for the best available location using version GBMRSP v2.0 of the response generator and version 2 of the GBM DRM database. 
Two sets of DRMs are generated, one for 8-channel (CTIME) data and one for 128-channel (CSPEC \& TTE) data. In case of relatively long duration GRBs multiple DRMs are used, which provide a new DRM  for every 2\degr\  slew of the \Fermi\  satellite  (RSP2).

\subsection{GRB Duration, Peak Flux and Fluence}

\begin{figure}[h]
\begin{center}
\epsscale{0.75}
\includegraphics[scale=0.60,angle=0.0,totalheight=0.85\textheight]{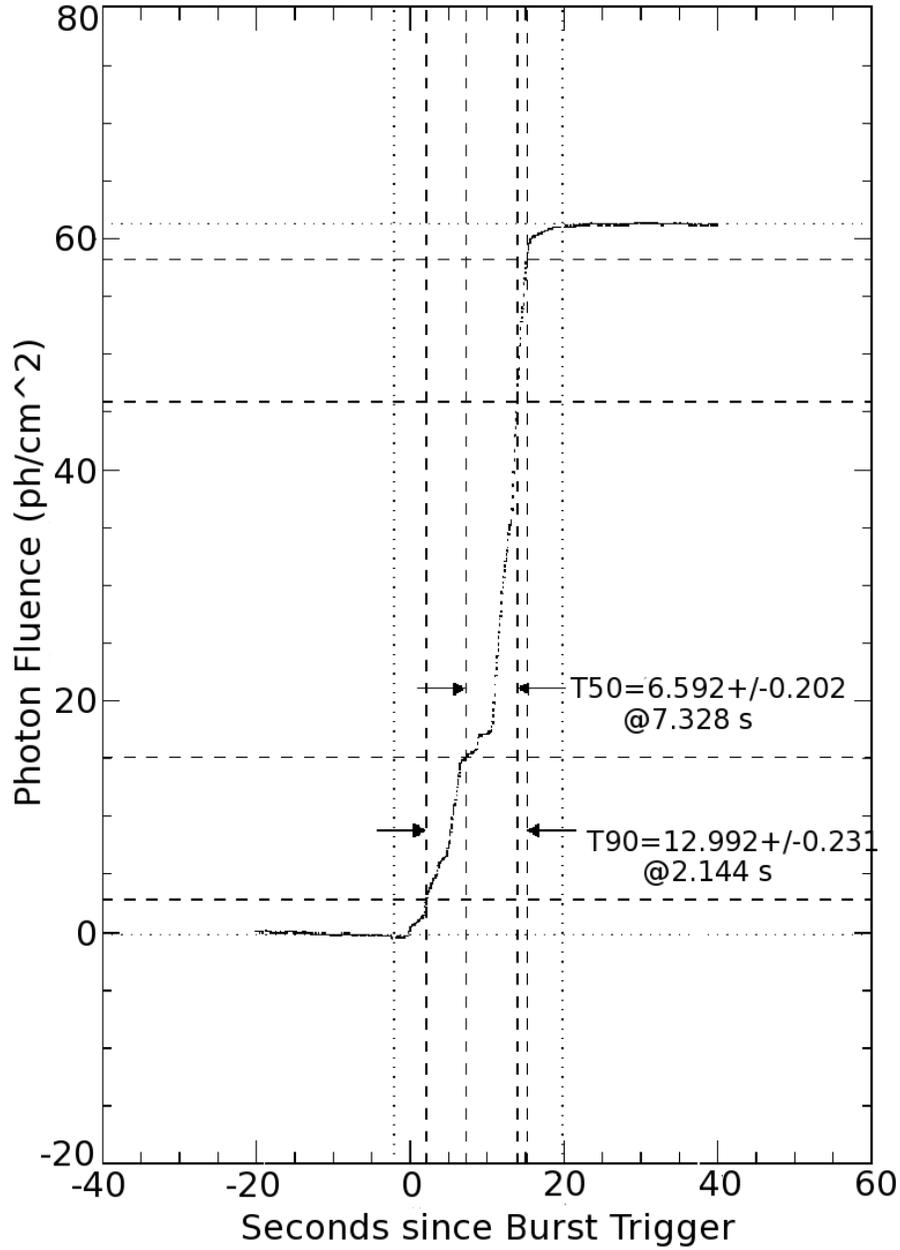}
\caption{\label{exampledur} The duration plot for GRB 131229A (bn131229277) is an example of the analysis for a GRB showing multiple pulses of different widths and amplitudes, some well separated and some overlapping. Data from NaI(Tl) detectors 9, 10 \& 11 were used. Horizontal dotted lines are drawn at 5\%, 25\%, 75\% and 95\% of the total fluence. Vertical dashed lines are drawn at the times corresponding to those same fluences, thereby defining the $T_{50}$ and $T_{90}$ intervals.}
\end{center}
\end{figure}

Analysis of GBM data products is fundamentally a process of hypothesis testing wherein trial source spectra and locations are converted to predicted detector count histograms, and these are statistically compared to the observed data. GBM uses a special burst spectroscopy software package called RMFIT\footnote{We used the spectral analysis package RMFIT, which was originally developed for time-resolved analysis of BATSE GRB data but has been adapted for GBM and other instruments with suitable FITS data formats. The software is available at the \Fermi\ Science Support Center: \url {http://fermi.gsfc.nasa.gov/ssc/data/analysis/user/} A tutorial is also available at \url{http://fermi.gsfc.nasa.gov/ssc/data/analysis/scitools/rmfit_tutorial.html}}. Here we report the duration, peak flux and fluence of each burst using an automatic batch fit routine implemented within RMFIT. 
Burst durations T$_{50}$ (T$_{90})$  are determined from the interval between the times where the burst has reached 25\% (5\%) and 75\% (95\%) of its maximum fluence, as illustrated by the horizontal and vertical dashed lines in figure \ref{exampledur}.
The burst durations T$_{50}$ and T$_{90}$ were computed in the 50-300 keV energy range.
This is primarily due to the fact that GRBs have their maximum spectral density in this energy range. In addition, this energy range makes it easier to compare the present results with those of the predecessor BATSE. It may be noted that GRB durations have been shown to have a power-law dependence on energy \citep{quin13}. The fluence for each burst was computed in two energy ranges: 50-300 keV and 10-1000 keV. Peak fluxes for each burst were computed in the same energy ranges and for three different timescales: 64 ms, 256 ms and 1024 ms. Since a relatively small number of bursts have detectable emission in the BGO detectors, only NaI(Tl) data were used for the catalog analysis.

\begin{figure}[h]
\includegraphics[scale=0.60,angle=0.0,totalheight=0.5\textheight]{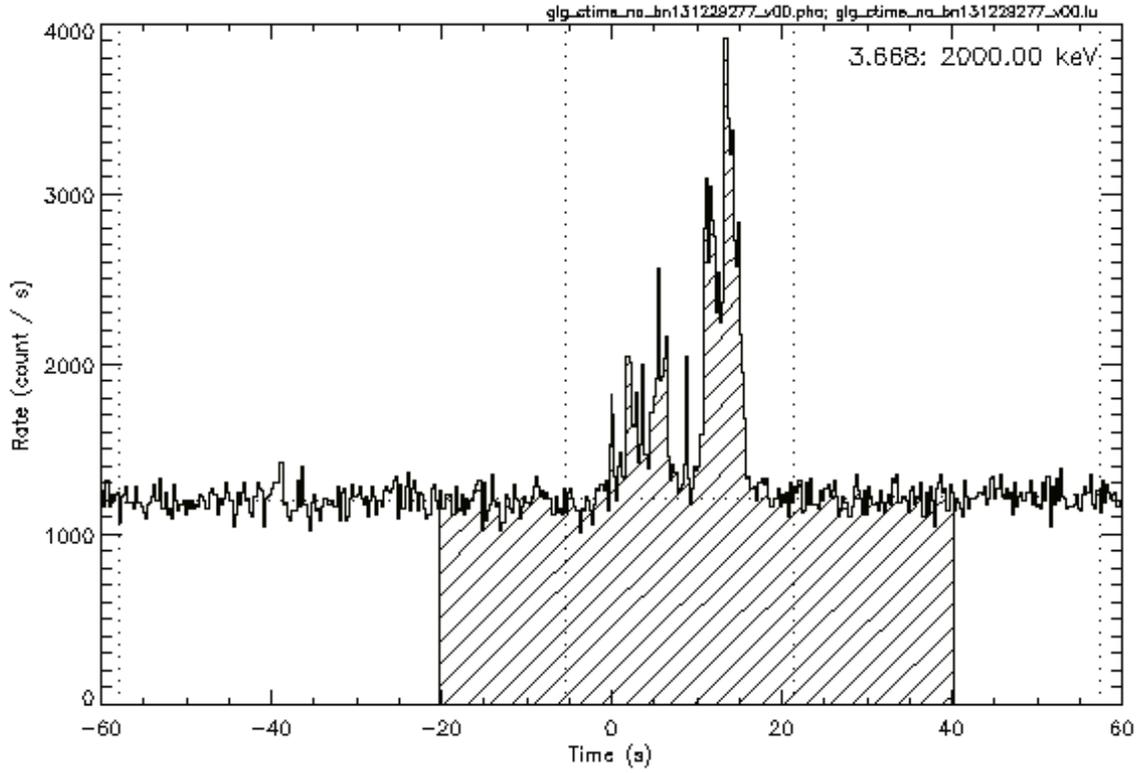}
\caption{\label{examplelc} CTIME lightcurve of GRB 131229A (bn131229277) with 0.256\ s temporal resolution in NaI(Tl) detector~10. Vertical dotted lines indicate the regions selected for fitting the background before and after the burst. The hatching defines the source region selected for the duration analysis. Note that the hatched region has good overlap with the background region before as well as after the burst.}
\end{figure}

For each burst, a set of NaI(Tl) detectors were chosen with low source angles (typically $< $ 60\degr) and
no apparent blockage by any other GBM detector, or LAT or the LAT radiators. For the majority of bursts the GBM CTIME data, which have 256 ms time resolution pre-trigger and 64 ms resolution post-trigger, were used. TTE data were used for bursts where at least one of the peak  fluxes occurs at or before the trigger time, which happens for many short bursts and a few longer ones. For each of the selected detector illuminated by a burst, source and background time intervals are then selected.  The source interval covers the burst emission time plus approximately equal intervals of background before and after the burst (generally at least $\sim$ 20\,s on either side for long GRBs while it is $\sim$ 5\,s or shorter for short GRBs). Background intervals are selected before and after the burst, about twice as wide as the burst duration and having a good overlap with the selected source interval before and after the burst as shown in figure \ref{examplelc}. 
Background regions before and after the burst are fit by a polynomial of up to 4$^{\rm th}$ order, separately for each detector. Depending on the background fluctuations, the lowest order polynomial that gives a good fit is chosen.
The background subtracted counts spectrum of each time bin in the source interval is fit to a model incident photon spectrum by folding the photon spectrum with the DRM and minimizing the CSTAT\footnote{\url{http://heasarc.nasa.gov/docs/xanadu/xspec/XspecManual.pdf}}. For the  T$_{90}$ analysis, the incident model spectrum is  ``Comptonized" (COMP) photon model:
\begin{displaymath}
f_{COMP}(E) = A \ \Bigl(\frac{E}{E_{piv}}\Bigr) ^{\alpha} \exp \Biggl[ -\frac{(\alpha+2) \ E}{E_{peak}}  \Biggr],
\end{displaymath}
The best fit parameters are: amplitude \emph{A}, the low energy spectral index $\alpha$ and peak energy $E_{peak}$ (the parameter $E_{piv}$ is fixed to 100 keV).

Figure \ref{examplelc} shows the light curve as measured by a single NaI(Tl) detector of a relatively long burst consisting of multiple emission periods, with the selected source and background intervals highlighted.  Figure \ref{exampledur} shows a plot of the integrated GRB fluence in the 50 -- 300~keV energy range derived from the model fitting for all time bins within the source interval. Several short duration plateaus seen in figure \ref{exampledur} are the time intervals where no burst emission is observed. This function is used to determine the T$_{50}$ (T$_{90}$) burst duration from the interval between the times where the burst has reached 25\% (5\%) and 75\% (95\%) of its total fluence, as illustrated by the horizontal dashed lines.

Peak fluxes and fluences are obtained in the same analysis, using the same choices of detector subset, source and background intervals and background model fits. The peak flux is computed for three different time intervals: 64 ms, 256 ms and 1.024 s in the energy range 10 -- 1000~keV and, for comparison purposes with the results presented in the BATSE catalog \citep{1998AIPC..428....3M}, in the 50 -- 300~keV energy range. The burst fluence is also determined in the same two energy ranges. The RMFIT analysis results presented above are stored in a BCAT fits file: glg\_bcat\_all\_bnyymmddttt\_vxx.fit with specified wildcards for the year (yy), month (mm), day (dd), fraction of a day (ttt) and version number (xx).

\section{GRB CATALOG RESULTS}
\label{CatRes}
The catalog results can be accessed electronically through the HEASARC browser interface at \url{http://heasarc.gsfc.nasa.gov/W3Browse/fermi/fermigbrst.html}. Here we provide tables that summarize selected parameters.

Table~\ref{main_table} lists the 1405 triggers of the first six years that were classified as GRBs. The GBM Trigger ID is shown along with a conventional GRB name as defined by the GRB-observing community.
Note that the entire table is consistent with the small change in the GRB naming convention that became effective on 2010 January 1 \citep{Barth09}: if for a given date no burst has been ``published'' previously, the first burst of the day observed by GBM includes the  `A' designation even if it is the only one for that day. For year 5 \& 6, only GBM triggered GRBs for which a Gamma-ray burst Coordinates Network (GCN) Circular was issued are assigned a GRB name. The criterion for issuing a GBM Circular is if a GRB was either detected by any other mission (as listed in the last column of table~\ref{main_table} ) or it generated an ARR to the \Fermi\ spacecraft  or the count rate in the 50-300 keV energy range  summed over the triggered detectors exceeded 1000 counts per second above the background.
The third column lists the trigger time in Universal Time (UT). The next four columns of Table~\ref{main_table} list the sky location and
associated statistical error\footnote{For GBM derived locations the statistical 1 $\sigma$ error is given. The GBM errors are not
symmetric and the given value is the average of the error ellipse.}, along with the instrument that determined the location.
The table also lists the GBM-derived location only if no higher-accuracy locations have been reported by any other instrument. If a higher-accuracy location is available its source is listed under the column  `Location Source' which lists only the name of the mission rather than the specific instrument onboard that mission (e.\,g., \Swift\  implies the locations are either from  \Swift-BAT or  \Swift-XRT or \Swift-UVOT); The errors on the GRB locations determined by other instruments are not necessarily 1 $\sigma$ values. For the GBM analysis, location accuracy better than a few tenths of a degree provides no added benefit because of significant systematic errors in GBM location \citep{Con15}.  Table ~\ref{main_table} also shows which algorithm was triggered along with its timescale and energy range. Note that the listed algorithm is the first one to exceed its threshold but it may not be the only one. The table also lists other instruments that detected the same GRB\footnote{This information was drawn from the GCN archive, accessible at \url{http://gcn.gsfc.nasa.gov/gcn3_archive.html}. A more complete list of detections is available at \url{http://www.ssl.berkeley.edu/ipn3/masterli.txt}.}. 
Finally, we identify the GBM GRBs for which an ARR was issued by the GBM FSW in the last column of table~\ref{main_table}.  We have a total of 93 GRBs (6.6\% of the total) followed by ARRs during the first 6 years of \Fermi,\ although the spacecraft might not have slewed in every case for technical reasons such as Earth limb constraints. The majority of these ARRs were due to high peak fluxes. In addition, there were 26 ARRs which were issued for non-GRB triggers because of the mis-classification by the GBM FSW.

The results of the duration analysis are shown in Tables~\ref{durations}, \ref{pf_fluence} \& \ref{pf_fluence_b}. The values of $T_{50}$ and $T_{90}$ in the 50--300~keV energy range are listed in Table~\ref{durations} along with their respective 1 $\sigma$ error estimates \citep{kosh96} and start times relative to the trigger time.  For a few GRBs the duration analysis could not be performed, either because the event was too weak or due to technical problems with the input data. Also, it may be noted that the duration estimates are only valid for the portion of the burst that is visible in GBM light curves summed over those NaI(Tl) detectors whose normals make less than 60\degr to the source. If the burst was partially occulted by Earth or had significant emission while GBM detectors were turned off in the SAA region, the ``true'' durations may be underestimated or overestimated, depending on the intensity and variability of the undetected burst emission. GRBs which triggered while \Fermi\ was close to SAA or the trigger is unusual in any other way, are indicated in table~\ref{main_table} with a footnote. For technical reasons it was not possible to do a single analysis of the unusually long GRB091024A \citep{2011A&A...528A..15G} and GRB130925A, so the analysis was done separately for the two triggered episodes. Similarly GRB130925A \citep{gre14,eva14} also had 3 emission episodes well separated in time, for which  GBM triggered on the first 2 episodes. These cases are also noted in the table~\ref{main_table}. The reader may note that for most GRBs the present analysis used data binned no finer than 64~ms, so the duration estimates (but not the errors) are quantized in units of 64~ms. However for a few extremely short events  TTE/CTTE data were binned with widths of  32~ms or even 16~ms in a few cases. 
As a part of the duration analysis, peak fluxes and fluences were computed in two different energy ranges. Table~\ref{pf_fluence} shows the values in 10-1000~keV and Table~\ref{pf_fluence_b} shows the values in 50--300~keV.
The analysis results for low fluence events are subject to large systematic errors primarily because they use 8-channel spectral data and should be used with caution. The fluence measurements in the spectroscopy catalog \citep{2014ApJS..211...12G}, which uses the 128-channel CSPEC or TTE data, are more reliable for such weak events.

\section{DISCUSSION}
\label{Disc}
\begin{landscape}
\begin{figure}
\begin{center}
\epsscale{0.95}
\plotone{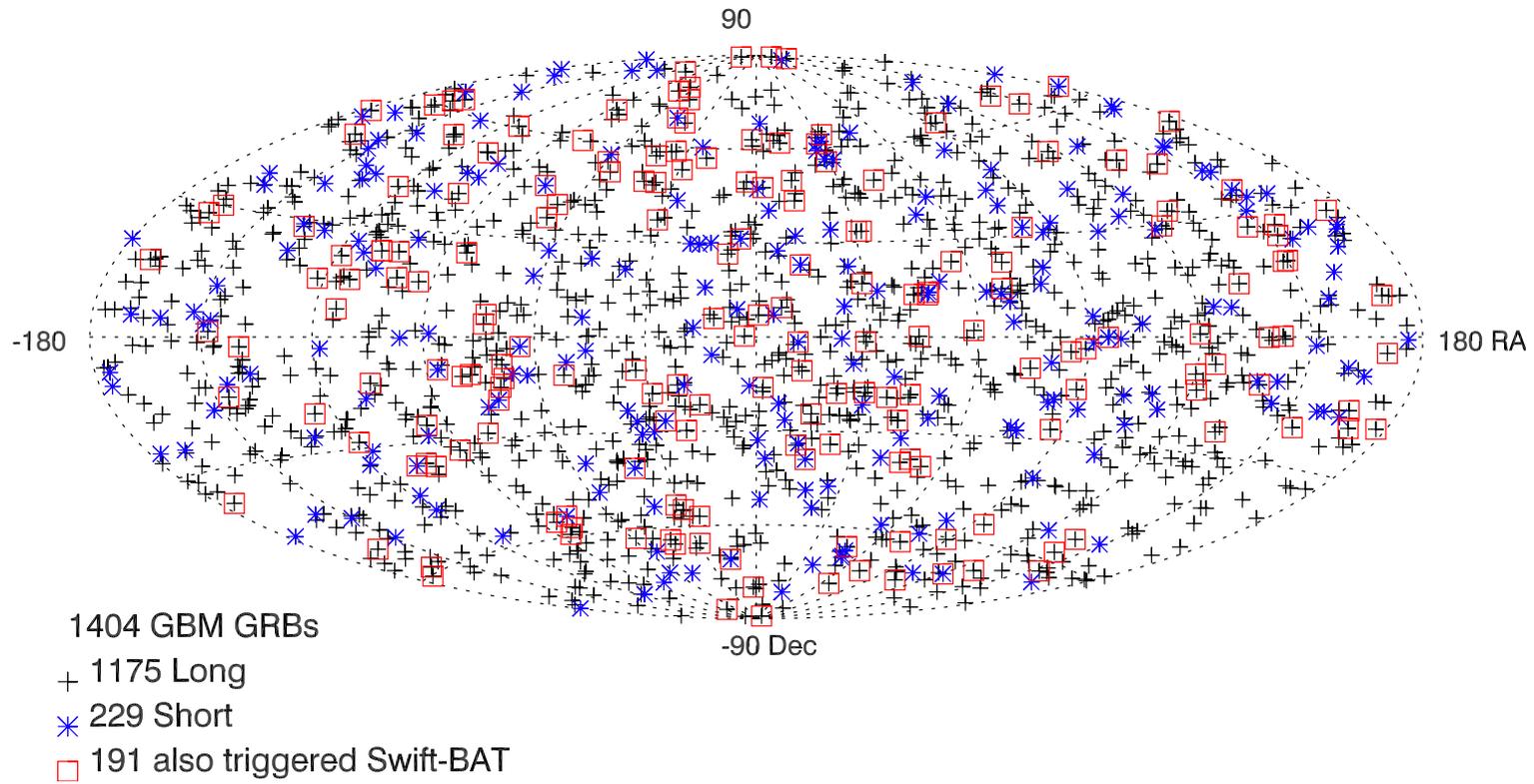}
\caption{\label{sky_dist} Sky distribution of GBM triggered GRBs in celestial coordinates. Crosses indicate long GRBs ($T_{90} > 2$~s); asterisks indicate short GRBs. Also shown are the GBM GRBs simultaneously detected by Swift (red squares)}
\end{center}
\end{figure}
\end{landscape}

Figure \ref{sky_dist} shows the sky distribution of GBM triggered GRBs in Galactic coordinates. Crosses indicate long
GRBs (T$_{90}\  >$ 2\,s); asterisks indicate short GRBs. Both the long and short GRB locations do not show any obvious anisotropy, consistent with an isotropic distribution of GRB arrival directions. Also shown are the locations of GRBs that triggered \Swift-BAT in coincidence with GBM. Many of these \Swift\ coincident GRBs also have redshifts estimated by detecting the optical afterglows by ground based telescopes.

\begin{figure}
\begin{center}
\epsscale{0.75}
\plotone{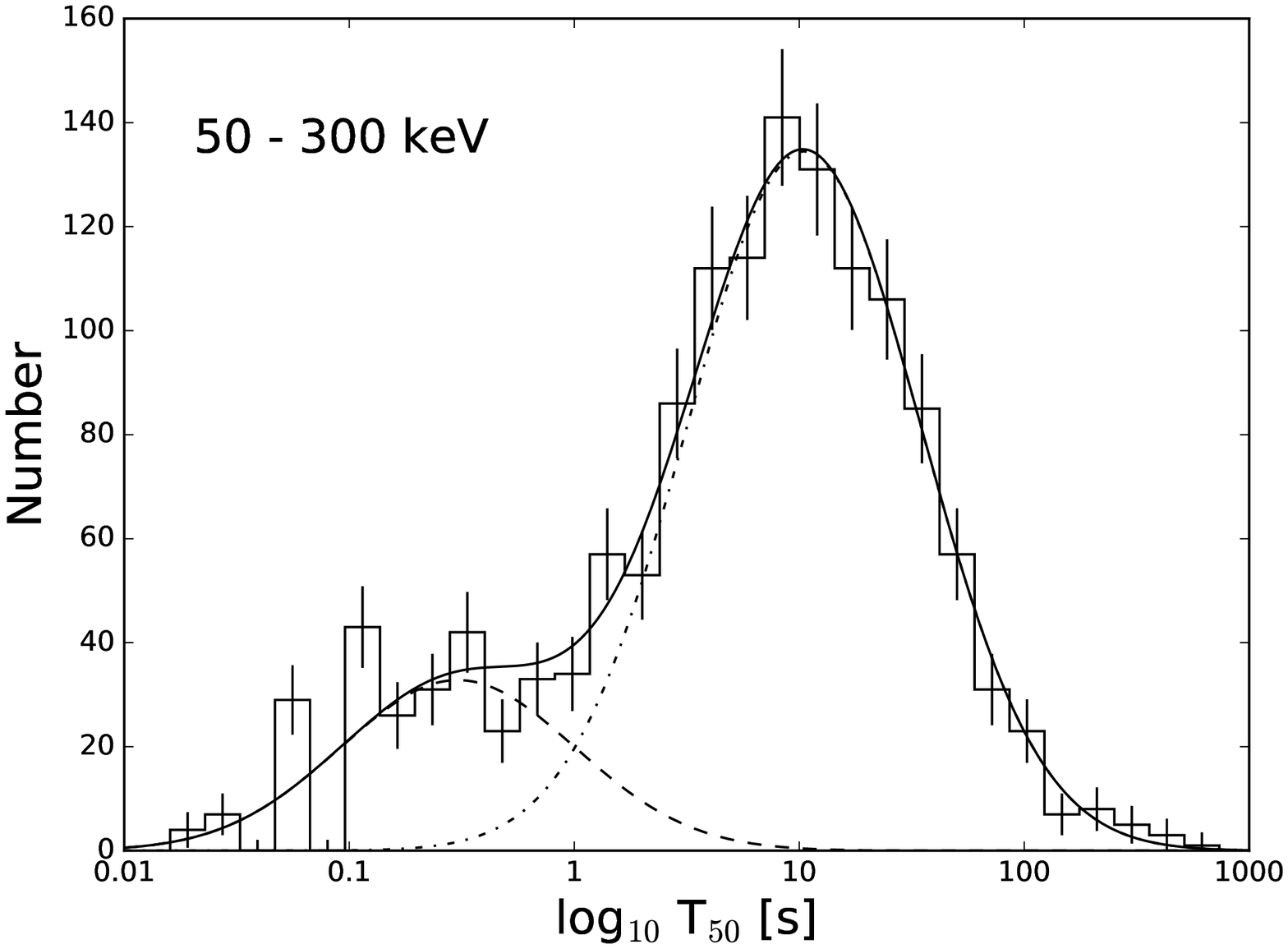}
\plotone{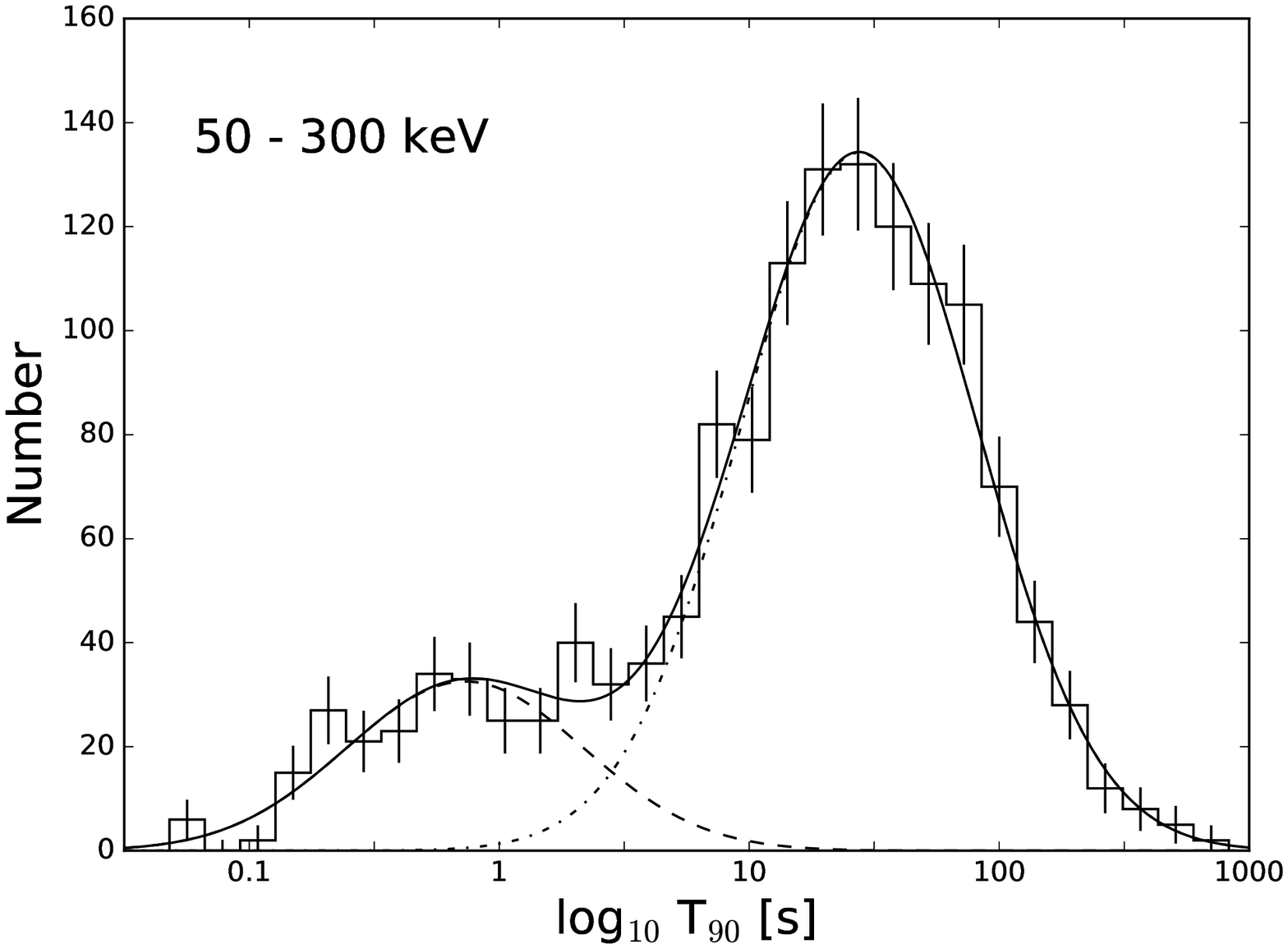}
 \caption{\label{dur_dist} Distribution of GRB durations in the 50--300~keV energy range. The upper plot shows $T_{50}$ and the lower plot shows $T_{90}$. Also shown are the lognormal fits separately to long and short GRBs (see text for details).}
\end{center}
 \end{figure}

\begin{figure}
\begin{center}
\epsscale{0.75}
\plotone{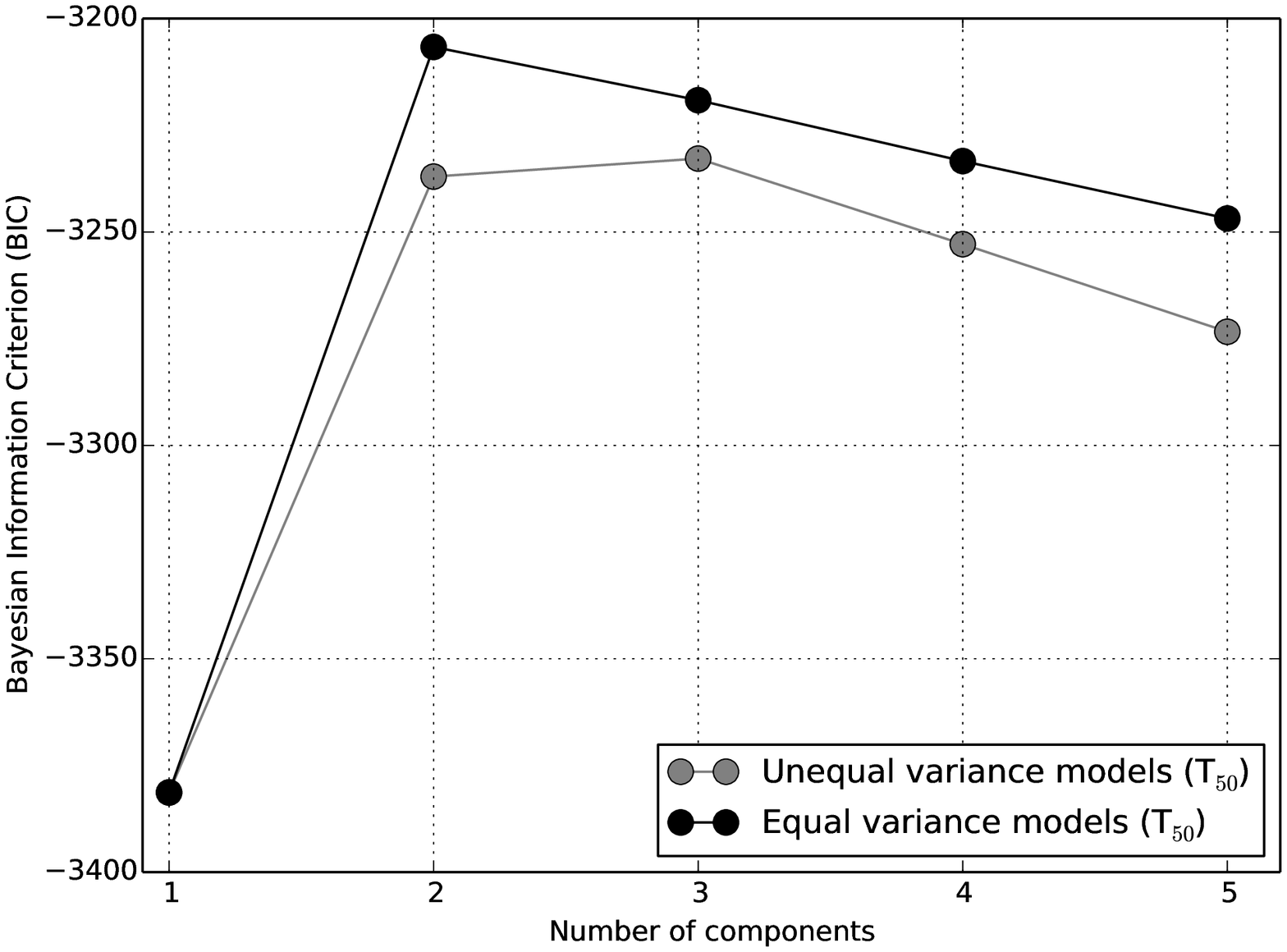}
\plotone{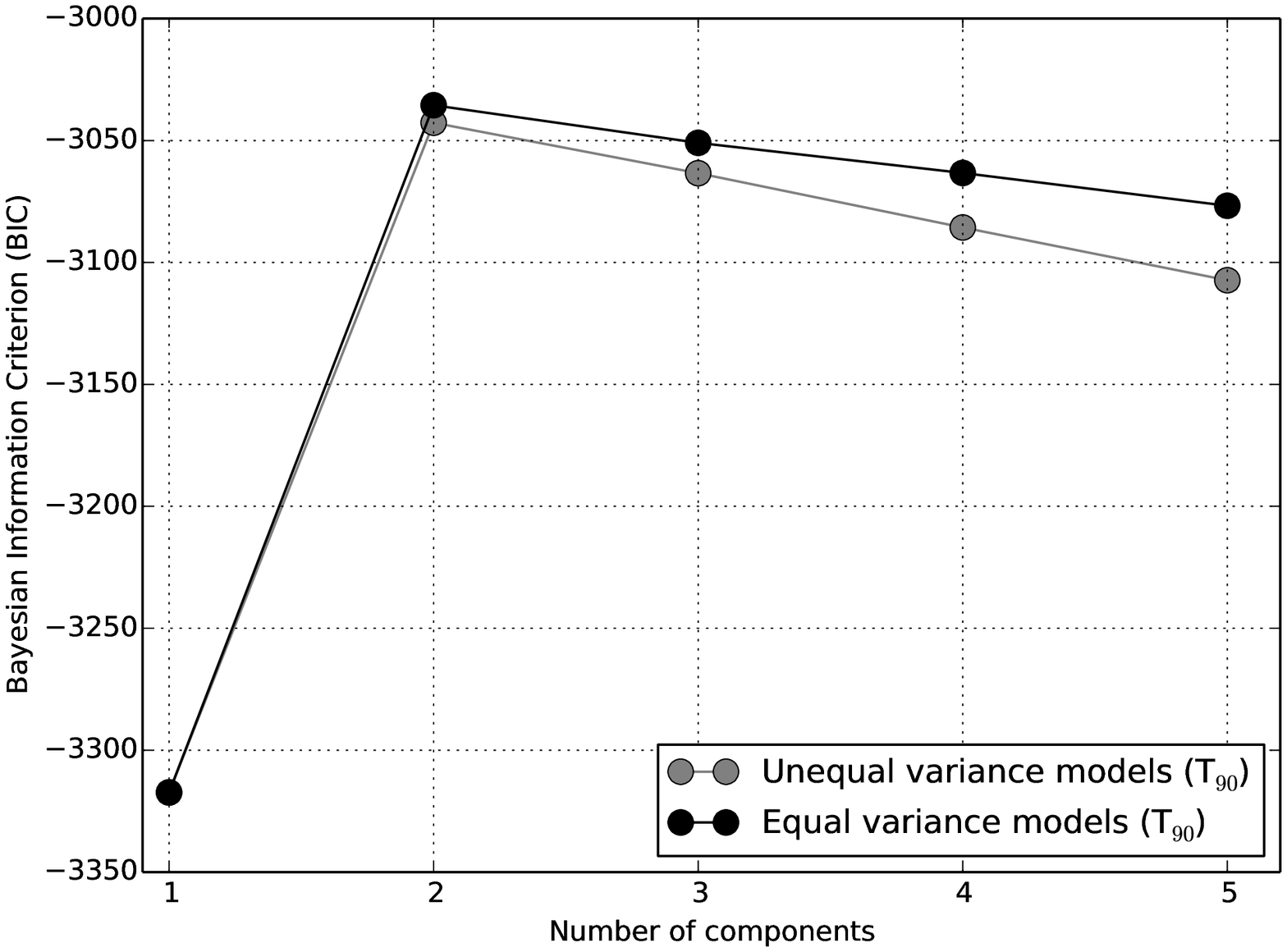}
 \caption{\label{bic_dist} 
 Figure showing the BIC values to establish the number of components that best
describe the GRB duration distributions. In the unequal variance model all parameters are left to
vary, while in the equal variance case, the variances of the components are
restricted to be equal among the components. The upper plot corresponds to $T_{50}$ and the lower plot corresponds to $T_{90}$ distributions. The peak at the number of components = 2 shows that the observed GRB duration distribution is best described by a 2 component model as shown in figure \ref{dur_dist}.
  }
\end{center}
 \end{figure}

The histograms of the logarithms of GBM triggered GRB durations ($T_{50}$ and $T_{90}$) are shown in Figure~\ref{dur_dist}. Using the conventional division between the short and long GRB classes ($T_{90}~ \leq$~ 2\,s and $T_{90} >$ 2\,s respectively) we find that during the first 6 years 
there were 229 short GRBs and 1175 long GRBs. The short and long GRBs, as defined by their $T_{90}$ in 50-300\,keV, possibly belong to two different classes \cite{CK93}. However from the $T_{90}$ distribution shown in figure \ref{dur_dist} the distinction seems to be less than obvious. 
{
There are also several claims in the literature on the existence of 3 types of GRBs based on multiple GRB parameters like duration, fluence, spectrum, spectral lag, peak-count rate etc., from BATSE sample \citep{muk98, hor06}, Swift sample \citep{ver10} as well as RHESSI sample \citep{ripa12}. The 3 groups are the familiar short-hard GRBs, long-soft GRBs and a soft-intermediate duration GRBs bridging the other two groups.
}
Hence we decided to independently assess the number of groups in GRB durations ( $T_{90}$ and $T_{50}$) as well as duration-hardness distributions by a model based clustering method using lognormal model components  \citep[`mclust' ][]{Fraley}.  This method uses a binning-independent maximum likelihood function with correction for the degrees of freedom, called Bayesian Information Criterion (BIC).  We find that both the log T$_{50}$ and log T$_{90}$ distributions are best described by two components with equal variance (see Fig \ref{bic_dist}). The difference in BIC values between two and three components is 12 for T$_{50}$ and $\sim$ 15 for T$_{90}$ (see Figure \ref{bic_dist}), both suggesting a strong preference for two groups of GRBs in the present sample. Figure \ref{dur_dist} also shows the lognormal fits separately to long and short GRBs. The variances of the lognormal components are constrained to be equal for each group. {
The goodness of fit is estimated by Kolmogorov-Smirnov test that yields a probability of 0.745 and 0.796 in favor of the null hypothesis (data and model are drawn from the same distribution) for T$_{90}$ and T$_{50}$ distributions respectively.}

In addition to  quantify the extent to which 2 groups are preferred compared to 3 populations statistically, we have carried out Monte Carlo simulations. First, we simulated $10^5$ instances of $T_{90}$ and $T_{50}$ with the best 2 group solution. We found that in 16 ($p=1.6\times10^{-4}$) and 24 ($p=2.4\times10^{-4}$) cases respectively the 3 group solution was preferred. To gauge the significance of the 3 group solutions we have simulated $10^4$ instances with the best 3 group solution for both $T_{90}$ and $T_{50}$. We found that a three group solution was only preferred in 3 and 41 cases respectively. This indicates that in the three group solution, the third group is not clearly distinguished.

\begin{figure}
\begin{center}
 \epsscale{0.5}
 \includegraphics[scale=0.60,angle=-90.0,totalheight=0.42\textheight]{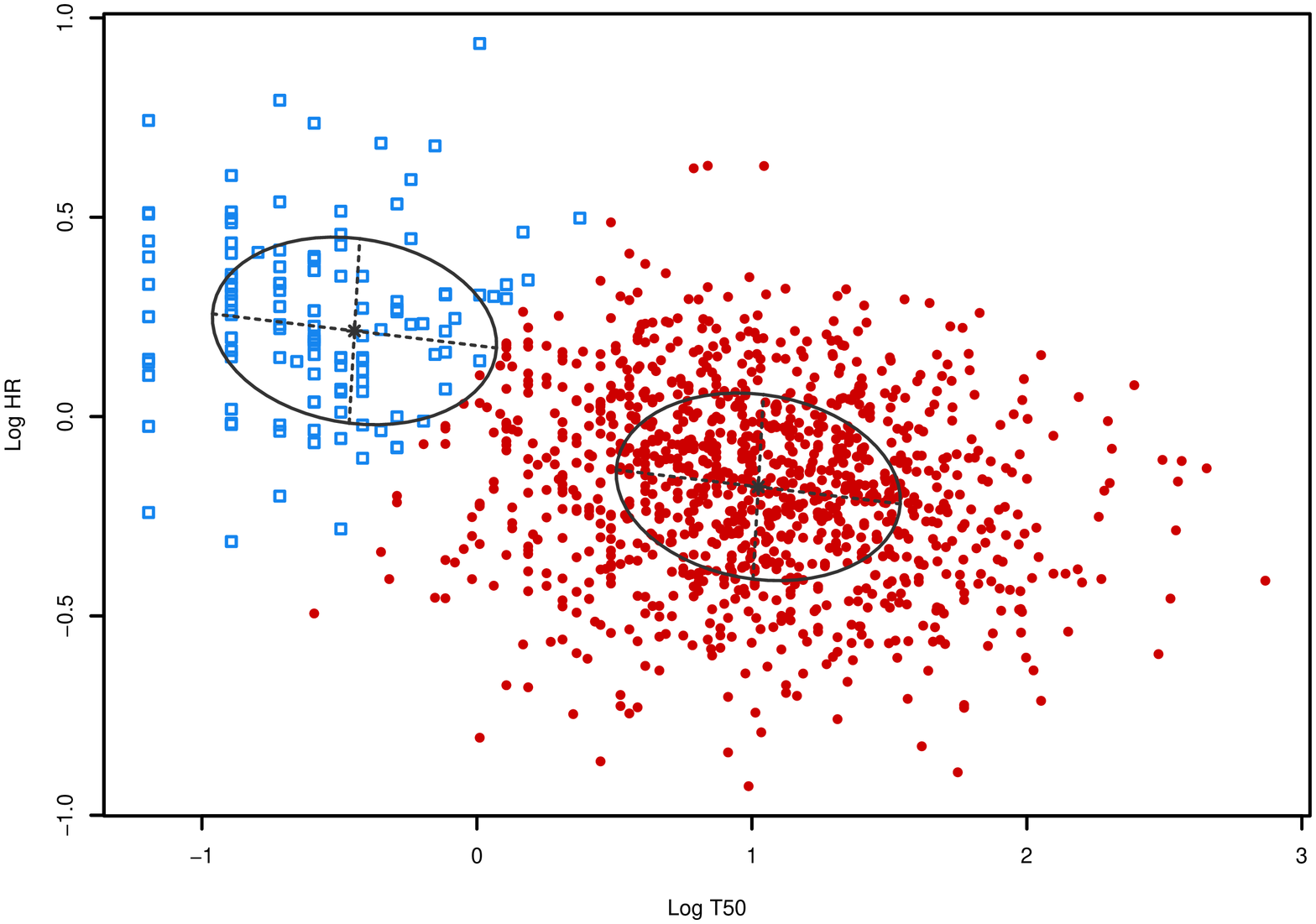}
\includegraphics[scale=0.60,angle=-90.0,totalheight=0.42\textheight]{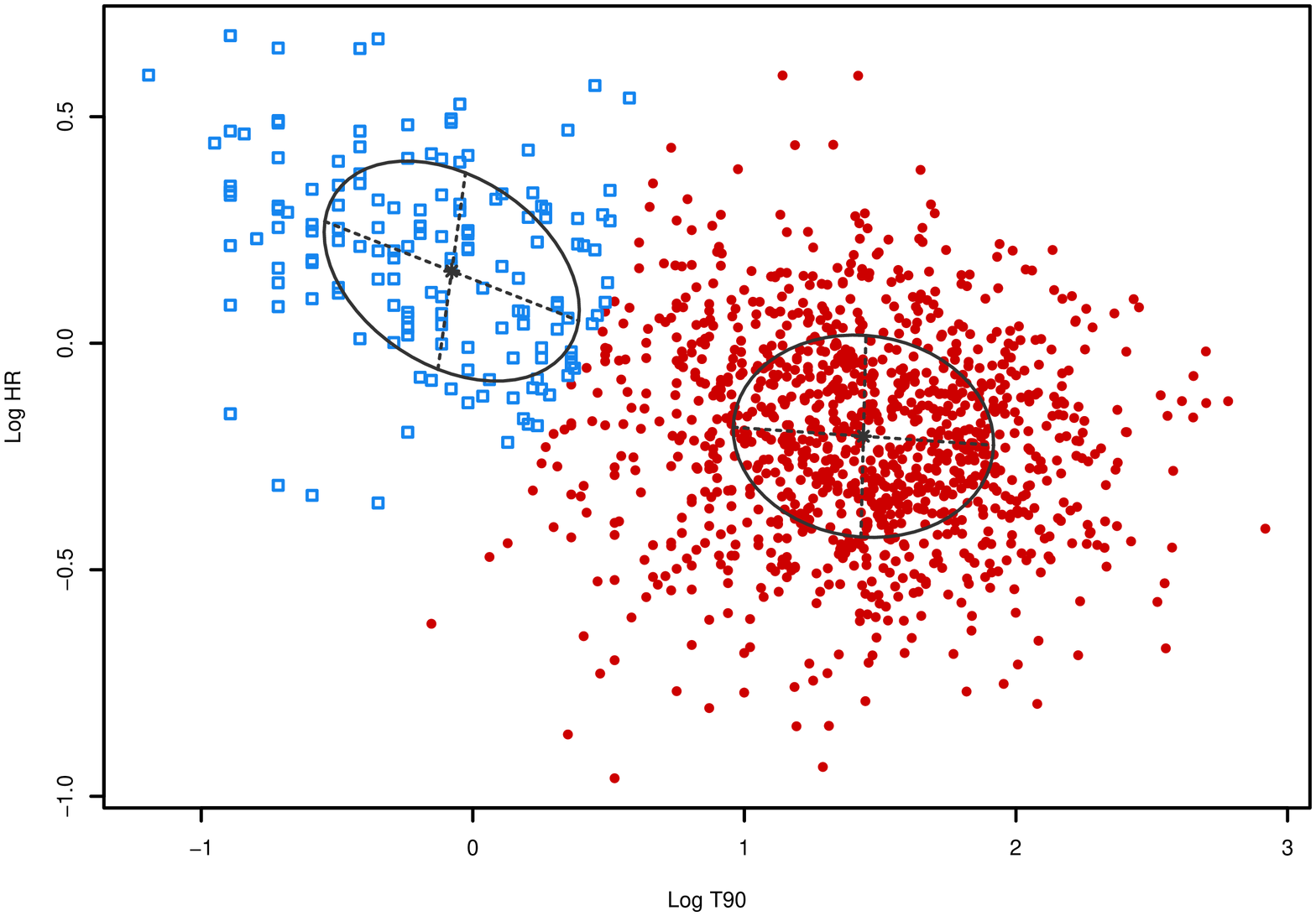}

\caption{\label{bic_hr} Classification based on the hardness-duration diagram. Here we show only GRBs with hardness errors less than the hardness itself. Colors indicate their group membership (red: on average short/hard, blue: on average long/soft). Ellipses show the best fitting multivariate gaussian models. In the T90-HR case (bottom) the best model has components with equal volume and shape (the major and minor axes of the ellipses are equal) but their orientation is not constrained. For T50-HR (top) the best model has similar properties as for T90-HR, only the orientation of the components is constrained to be the same (see Figure \ref{bic_tt} for BIC values of different models).}
\end{center}
\end{figure}

\begin{figure}
\begin{center}
 \epsscale{0.75}
\plotone{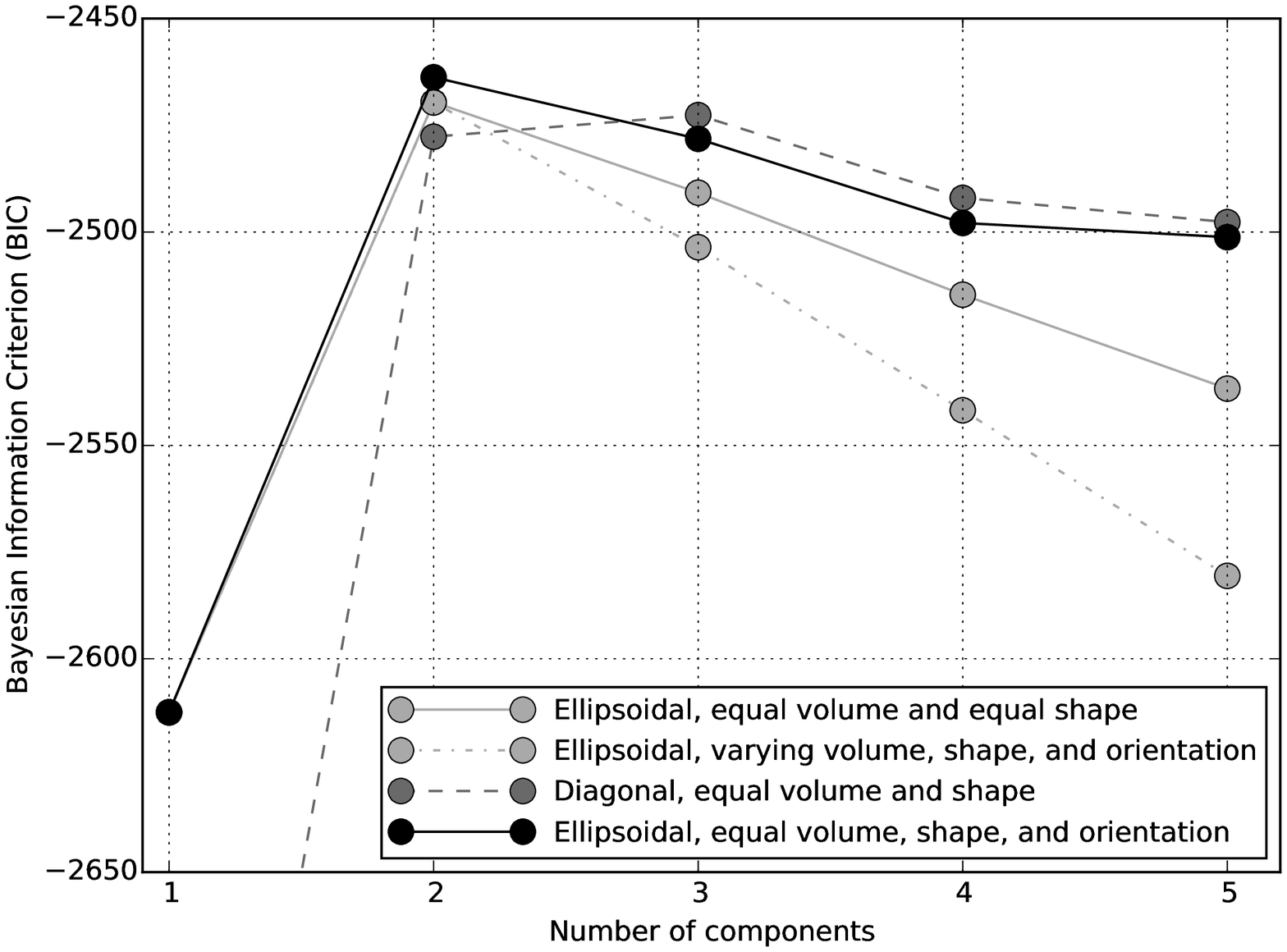}
\plotone{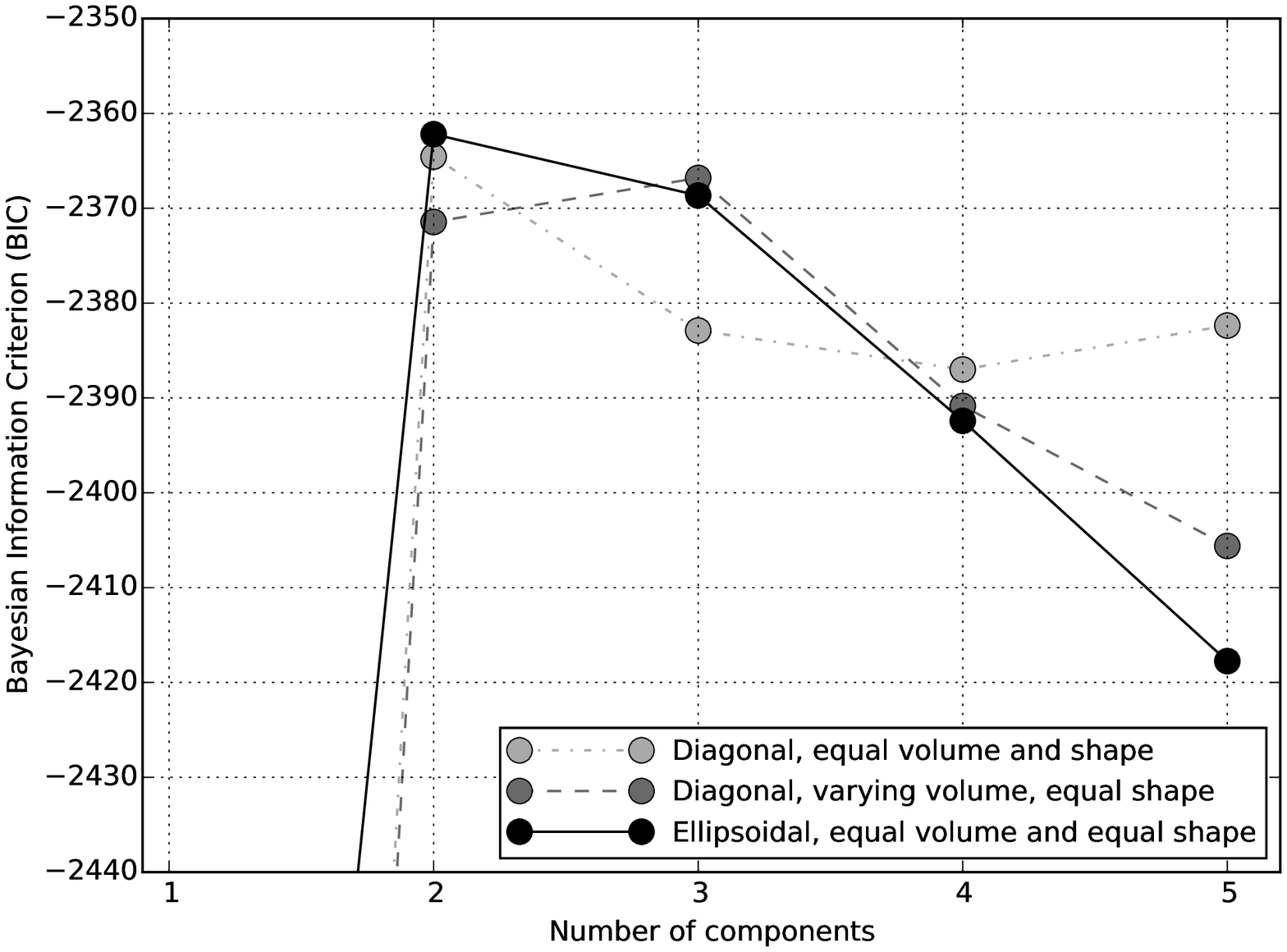}
\caption{\label{bic_tt}BIC values for the hardness-duration data [T90-HR (bottom) and T50-HR (top)] for the relevant bi-variate normal component models. For clarity, we only show the best faring models. See Figure \ref{bic_hr} for the realization of the best models.}
\end{center}
\end{figure}

Because the GRB groups have less overlap if we consider more than one
dimension, we consider the hardness in addition to duration (discussed later in this section). We use only those bursts
that have hardness errors less than the hardness value. We end up with a sample size of 1222.
Using  the BIC analysis again, we find that similar to the one dimensional
distribution of durations, the duration-hardness data are also best described by two groups
(see figure \ref{bic_hr}).  In figure \ref{bic_hr} the ellipses mark the one sigma contours of the two dimensional
Gaussians. They encompass $\approx 0.39$ of the volume of the
individual components (this is analogous to the 0.68 fraction marking
the one sigma region in the one dimensional Gaussian case). We find that $40.1\%$ and
$39.8\%$ GRBs are contained within the ellipses for the short and long cases respectively, consistent with the
expectations.
The difference between the best 2 and 3 group solutions is 4.6
and 8.9 for the $T_{90}-HR$ (bottom) and $T_{50}-HR$ (top) distributions respectively. In the parlance of the
BIC values, these constitute strong evidence {\it for} the 2 group solution. In
order to have a more quantitative assessment we, again, have simulated
distributions with the best 2 group solution 
and found 0 cases in $10^3$ trials ($p<10^{-3}$) and 3
($p=3\times10^{-3}$) cases where the 3 group solution was preferred in the
$T_{90}-HR$ and $T_{50}-HR$ distributions respectively.  In the case of
$T_{90}-HR$ distribution the best model has ellipsoidal components with equal volumes and
shapes, but their orientation is free to vary  (figure \ref{bic_tt} black solid line).
For $T_{50}-HR$ distribution the best model also has ellipsoidal components while their volumes, shapes
and orientations are constrained to be equal (figure \ref{bic_tt} black solid line).  In short, the model-based clustering
method applied to hardness ratios and durations unveils only
two clusters as the best solution: classical short/hard and
long/soft groups consistent with a similar analysis carried out on the RHESSI data \citep{ripa12}.  

Using an entirely independent approach to estimate the number of populations that can exist in the observer-frame $T_{90}$ distribution, we employ a Bayesian Dirichlet mixture model composed of Gaussians \citep[see][]{gre12}. This enables us to ask the question how many sub-populations exist rather than asking whether three populations fit the data better than two populations. We use an approach similar to that followed by \citet{cha07} except that we adopt a hierarchical Bayesian approach which allows us to leave the concentration as a free parameter in the model. The model returns posterior distributions of the probability for each sub-population that is found. We find that there are two significant Gaussian sub-populations with 95\% highest density intervals for their existence covering p=0.77-0.84 and p=0.15-0.23 corresponding to Gaussian means of 27.5 s and  0.79 s respectively for long and bursts. Since the entire Dirichlet must sum to p=1, this leaves little room for a third sub-population. In fact, the next highest probability for an additional subpopulation is 0.001. Additionally, we checked that using Student-t distributions to model the sub-populations, {\it i.e.}, seeing if non-normality or outliers changed the distributions. We found similar results with the Student-t distributions converging to Gaussians. We can therefore conclude that there exist only two sub-populations in the T90 distribution of GBM detected GRBs. 

For a comparison with BATSE distribution of GRB durations, we have performed the same classification on the current BATSE catalog of 2041
GRBs consisting of 500 short and 1541 long GRBs \footnote{\url{http://gammaray.nsstc.nasa.gov/batse/grb/catalog/current/index.html}}.
For GBM GRB sample, the best
fitting model is two components with equal variances, while for BATSE sample it is the
two component with {\it un}equal variances. To compare similar quantities, we
forced the unequal variance model for GBM (which gives only a marginally worse fit) and compared it to the BATSE models.
The mean $T_{90}$ for the short bursts for BATSE vs. GBM are $T_{\rm mean,short}^{\rm
BATSE}=0.85$ s, $T_{\rm mean,short}^{\rm GBM}=0.82$ s for the long,
similarly BATSE vs. GBM $T_{\rm mean,long}^{\rm BATSE}=35.6$ s,   $T_{\rm
mean,long}^{\rm GBM}=28.3$ s. The mean durations of the short and long GRBs are consistent with those estimated above by the Bayesian analysis. It may be noted that the mean durations of the GBM detected long GRBs is smaller than those of BATSE detected long GRBs. This could be an indication of the well known tip-of-the-iceberg effect resulting from the higher sensitivity of BATSE detectors which are 16 times larger than the GBM detectors. 

The standard deviations of $\log_{10} (T_{90}/{\rm s})$ for short bursts in the
GBM sample is 0.51 while for BATSE it is 0.64. The same quantity for long GRBs
in the GBM sample is 0.64 while for BATSE it is 0.42. The larger difference between the widths of long and short GRBs of the BATSE sample explains the preference for the model with unequal variances. However the reason for unequal widths is not clear.

Since there are several changes and interruptions in GBM triggers and trigger algorithms during this catalog period (see table \ref{trigger_modification_history}) it will not be strictly accurate to get the burst rate by dividing the number of bursts by the total duration of 6 years. Instead we derived the average GRB rates by fitting the integral interval distributions of the burst trigger times to an exponential distribution function. The estimated daily rate of short GRBs  is (0.137 $\pm$  0.009) while that of long GRBs is 0.602 $\pm$ 0.018.
The fraction of short GRBs is  0.207 $\pm$ 0.015. Selecting periods where all three BATSE trigger algorithm were set to the same value (e.g. a threshold of $5.5 \sigma$ from 1992 September 14 to 1994 September 19 and from 1996 August 29 to the end of the mission) the observed fraction of short GRBs is 24\% \citep{1999ApJS..122..465P}. The shortest and longest time intervals between consecutive short bursts are 1.34h and 59 days, respectively, while the average interval is 7.28 days. Similarly  the shortest and longest time intervals between consecutive long bursts are 631s  and 11.8 days (with no interruptions) while the average interval is 1.66 days.
As already claimed in the first catalog we ascribe the lower fraction of short GRBs observed with GBM (20.7\%) compared to BATSE (24\%) not to a deficit of short events but rather to an excess of long events detected by GBM's longer timescale trigger algorithms (see Section \ref{trigger_statistics}). {
The average GBM GRB trigger rate of $\sim$242 $\pm$ 6.5 bursts/year is comparable to the BATSE rate of ~300 bursts/year, in spite of the large difference in area between the BATSE and GBM detectors.  The BATSE detectors were a factor of 16 larger in area, as mentioned before, resulting in a difference of a factor of 4 in sensitivity.  However, the logN-logP curve is much flatter near the BATSE threshold than the -3/2 power law seen at higher intensities.  Also, the GBM trigger thresholds is set at 4.5 sigma above threshold, while for BATSE it was set at 5.5 sigma.  The higher setting was needed to reject BATSE triggers due to fluctuations in the flux from Cygnus X-1.  Although GBM employs 12 detectors compared to 8 for BATSE, the sky coverage is about the same due to obstructions by the LAT on Fermi, as opposed to clear views obtained on Compton Gamma-ray Observatory (\CGRO).  Most significantly,  GBM triggers on 5 timescales from 16 ms to 4096 ms, at two different phases, whereas BATSE triggered only on 3 timescales 64, 256, and 1024 ms. Pre-launch predictions of the GBM burst rate were approximately 200 bursts/year based on scaling from BATSE, and assuming the same trigger algorithms.  This is, within statistics, the observed rate of GBM triggers that would have triggered on one of the BATSE timescales.  The additional $\sim$42 bursts/year arise mostly from the triggers on the new 4096 ms timescale.  

Since the orbits of Fermi and GRO are similar, GBM and BATSE should have similar celestial sensitivity maps.  However, sensitivity sky maps are not maintained for GBM since the isotropy of gamma-ray bursts is no longer controversial.
}
Furthermore GBM slightly favors triggering on long GRBs,  since the thresholds for the~64 ms timescales are higher ($5.0 \sigma$, see Table \ref{trigger:criteria:history}) than for 256 \& 1024~ms (both $4.5 \sigma$)\footnote{It may be noted that there were also times when BATSE triggers did not use the same threshold for all 3 timescales \citep[see Table 1 in][]{1999ApJS..122..465P}. see \url{http://heasarc.gsfc.nasa.gov/W3Browse/all/batsegrb.html}}.

\begin{figure}
\begin{center}
\epsscale{0.7}
\plotone{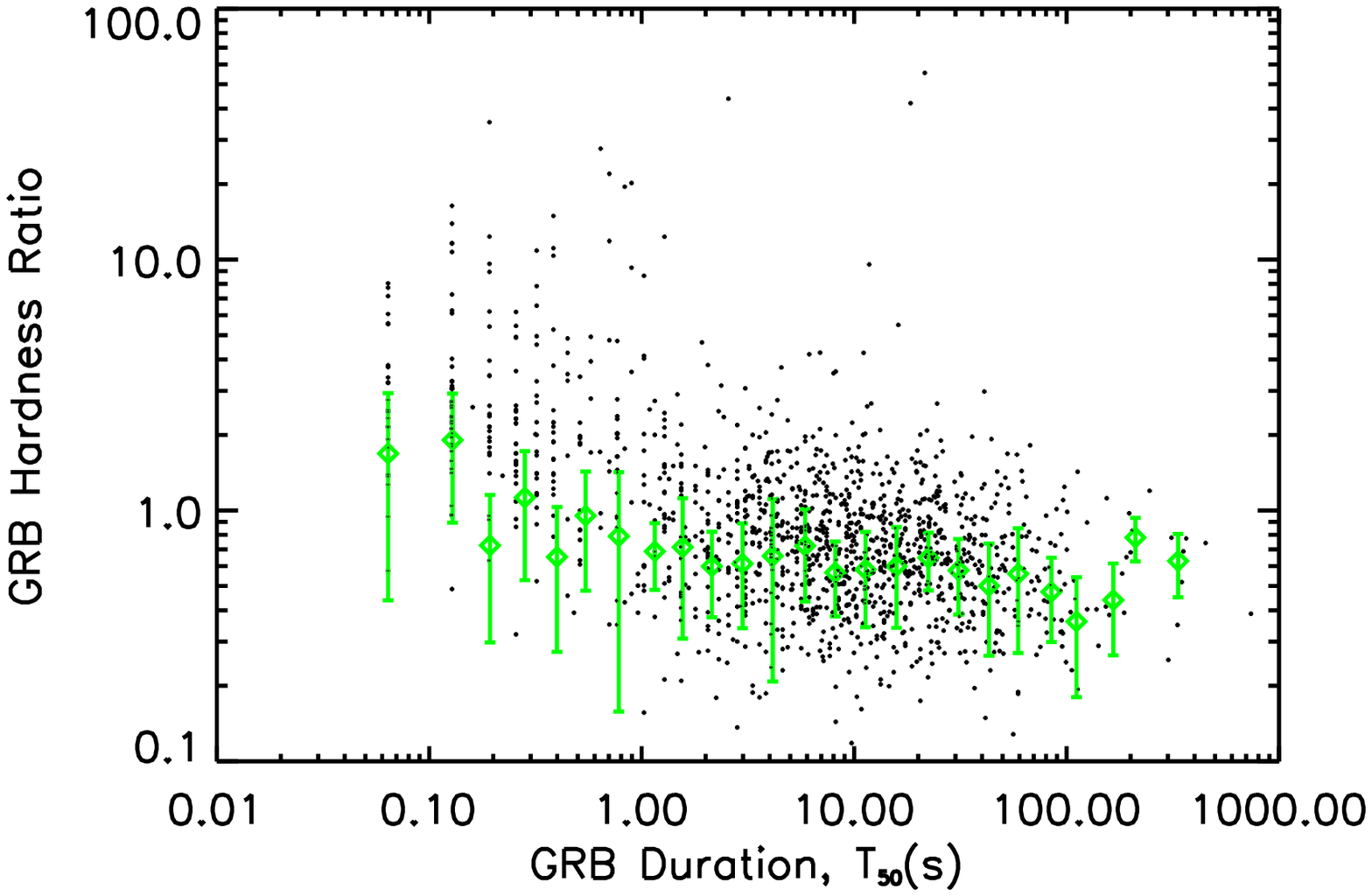}
\plotone{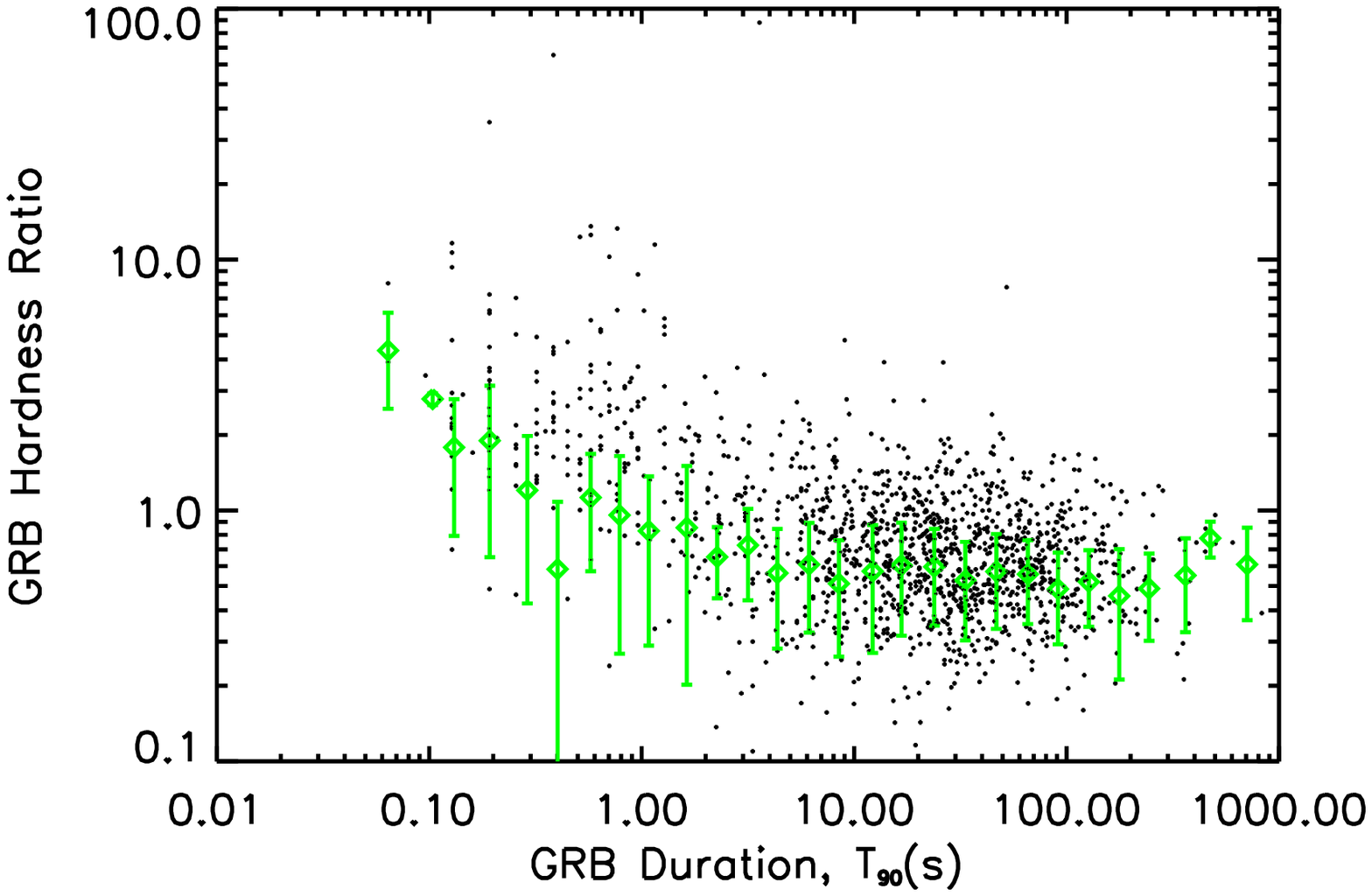}
\caption{\label{hardness_vs_dur} Logarithmic scatter plots of spectral hardness vs.\ duration are shown for the two duration measures $T_{50}$ (upper plot) and $T_{90}$ (lower plot). The spectral hardness was computed from the duration analysis results by summing the deconvolved counts in each detector and time bin in two energy bands (10 -- 50~keV and 50 -- 300~keV), and further summing each quantity
in time over the $T_{50}$ and $T_{90}$ intervals.  The
hardness ratio was calculated separately for each detector as the ratio of the flux density
and finally averaged over detectors. The error bars for individual bursts are suppressed for clarity.   1376 $T_{90}$  hardness ratios and 1364  $T_{50}$ hardness ratios out of a total sample of 1403 GRBs have been estimated and plotted here. The rest are too weak to compute hardness ratios and hence ignored. 
 The short and long GRBs were further divided in to equal logarithmic duration sub-groups. The green points with errors are the average values of hardness ratios weighted with inverse of their errors over GRBs that fall under each group. The anti-correlation of spectral hardness with burst duration is evident.}
\end{center}
\end{figure}

To characterize the dependence of burst spectral hardness on duration, we computed the hardness of each GRB as the ratio of burst fluence during the $T_{50}$ or $T_{90}$ intervals in the energy band  50 -- 300~keV to that in the 10 -- 50~keV band. In this analysis, the hardness was derived from the time-resolved spectral fits for each GRB by using photon model fit parameters that are a by-product of the duration analysis.  Figure~\ref{hardness_vs_dur} shows scatter plots of hardness vs.\  $T_{50}$  and $T_{90}$  durations,  showing that the GBM data also exhibit the anti-correlation of spectral hardness with duration as known from BATSE data \citep{CK93}. However the individual data points are somewhat misleading, as the high hardness ratios tend to have large statistical errors, which are not plotted for the sake of clarity. Hence we also computed weighted mean hardness ratios for GRBs in equal bins of logarithmic durations\footnote{
Note that these are NOT the same as weighted means of log(HR)}. These are shown in green with their corresponding weighted errors. The green points indicate a gradual hardening of the GRBs as the durations fall below $T_{90}$ = 2\,s, which is in contrast to PHEBUS\footnote{PHEBUS was a Soviet-French gamma-ray burst experiment onboard GRANAT satellite launched before \CGRO.} and BATSE observations, that showed a more discrete change in hardness at $T_{90}$ = 2\,s \citep{1992AIPC..265..304D,2003AIPC..662..248P}. However we do not see any positive correlation of hardness ratio with duration for long GRBs as reported before \citep{1996ApJ...471L..27D}. Instead, we see a weak anti-correlation (correlation coefficient of -0.17) of hardness ratio with duration of long GRBs in our 6 year sample.

\begin{figure}
\begin{center}
 \epsscale{0.9}
\plotone{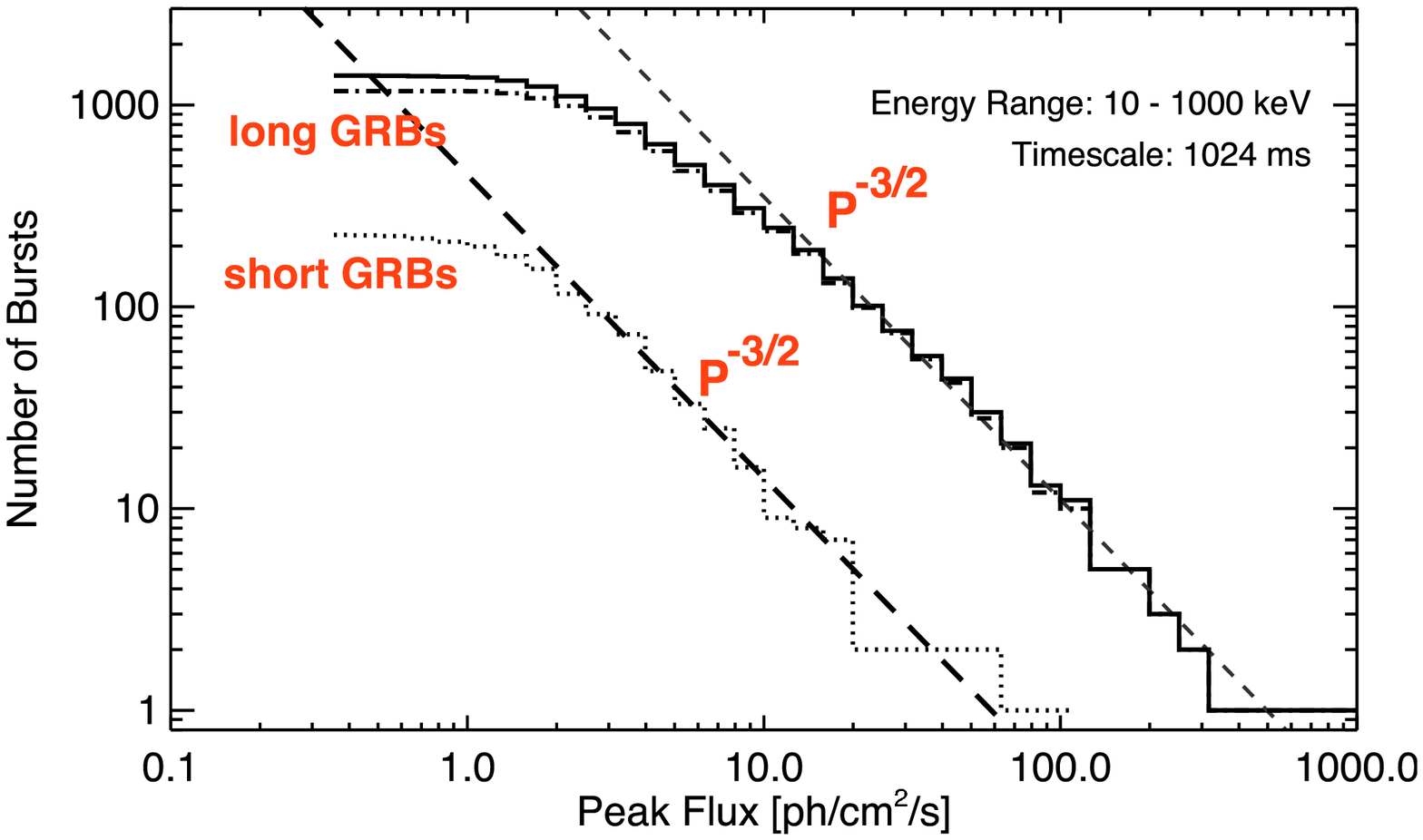}
\plotone{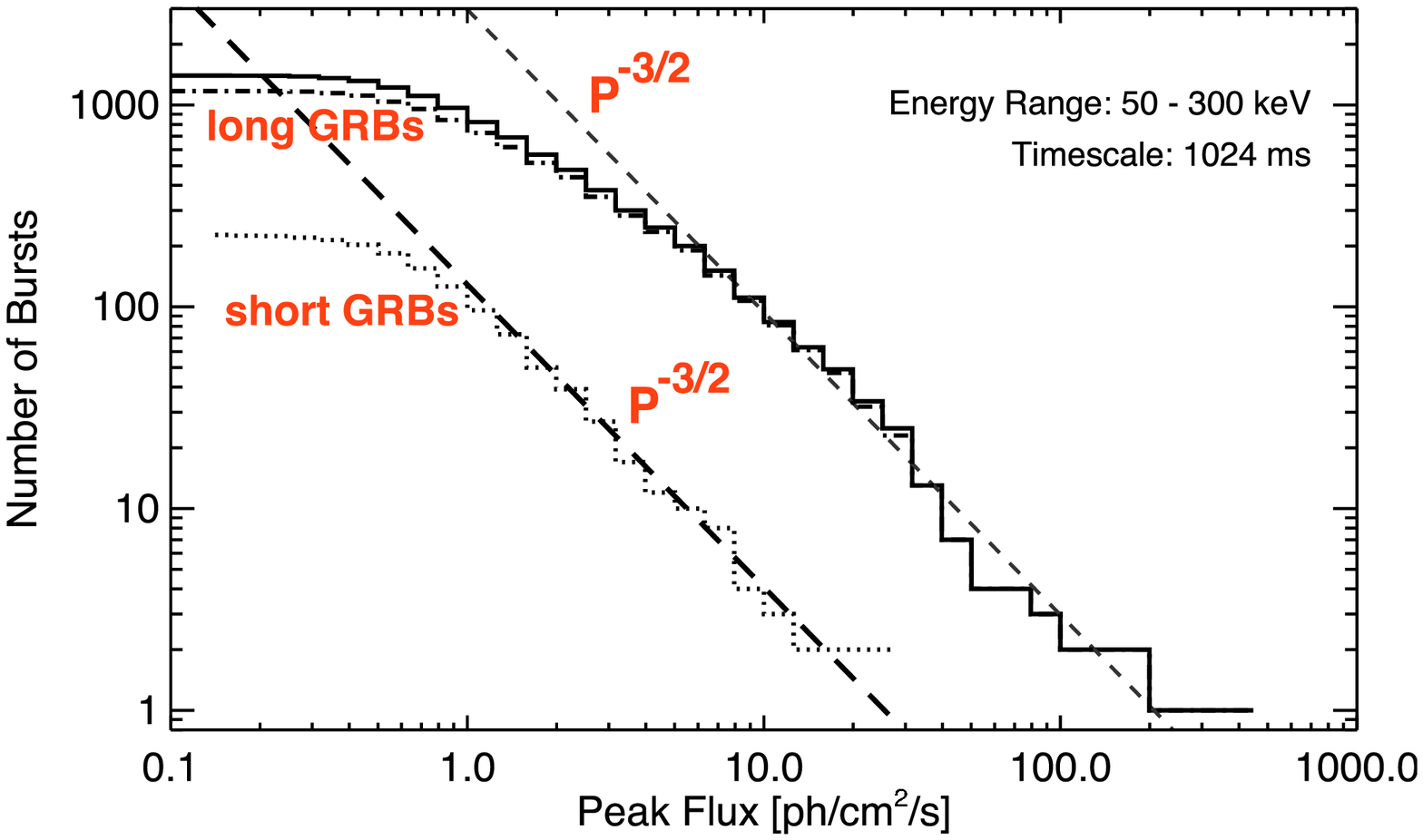}
\caption{\label{pflx_fig} Integral distribution of GRB peak flux on the 1.024~s timescale. Energy ranges are 10 -- 1000~keV (upper plot) and 50 -- 300~keV (lower plot). Distributions are shown for the total sample (solid histogram), short GRBs (dots) and long GRBs (dash-dots), using $T_{90} = 2$~s as the distinguishing criterion. In each plot a power law with a slope of $-3 / 2$ (dashed line) is drawn to guide the eye.}
\end{center}
\end{figure}

\begin{figure}
\begin{center}
 \epsscale{0.95}
\plotone{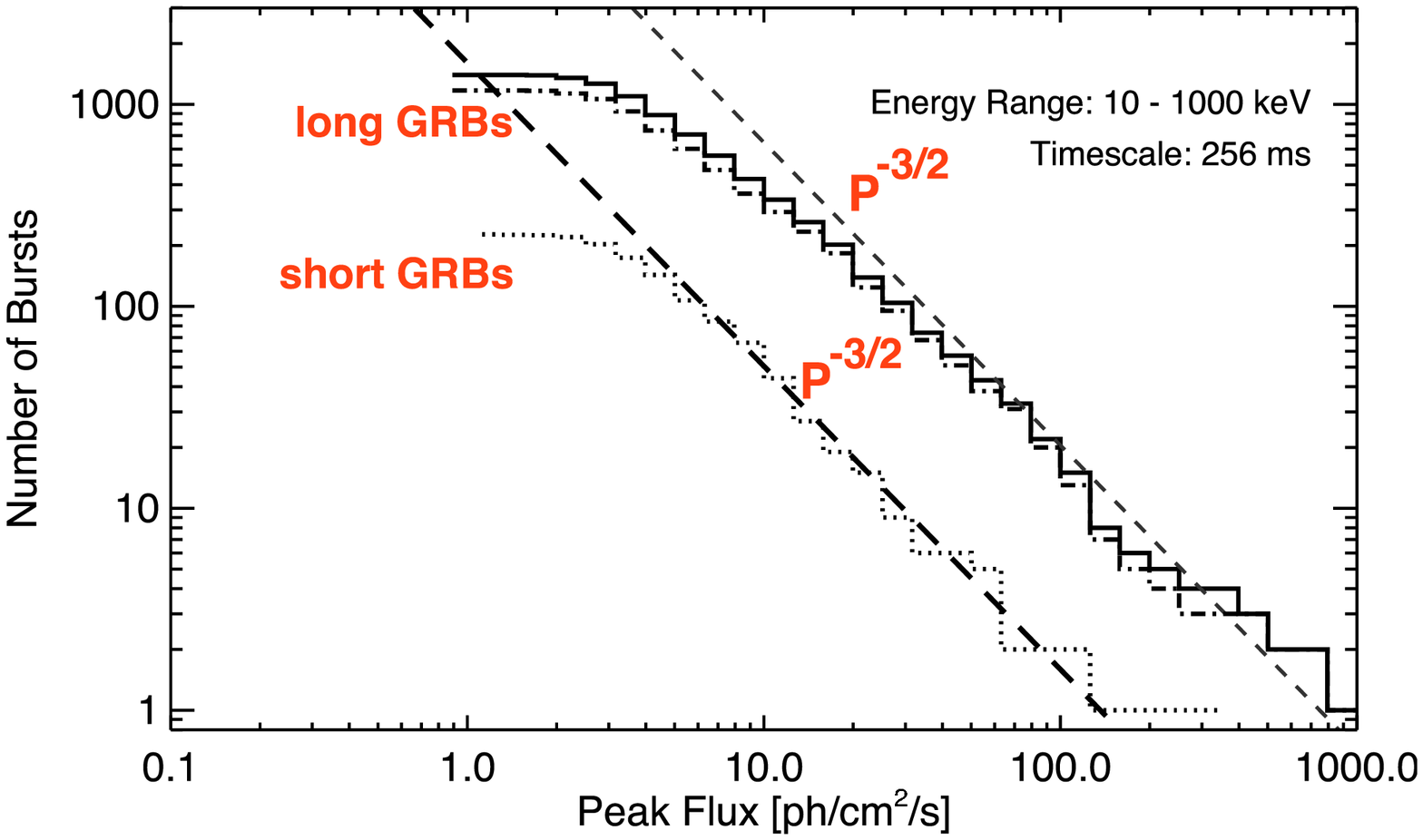}
\plotone{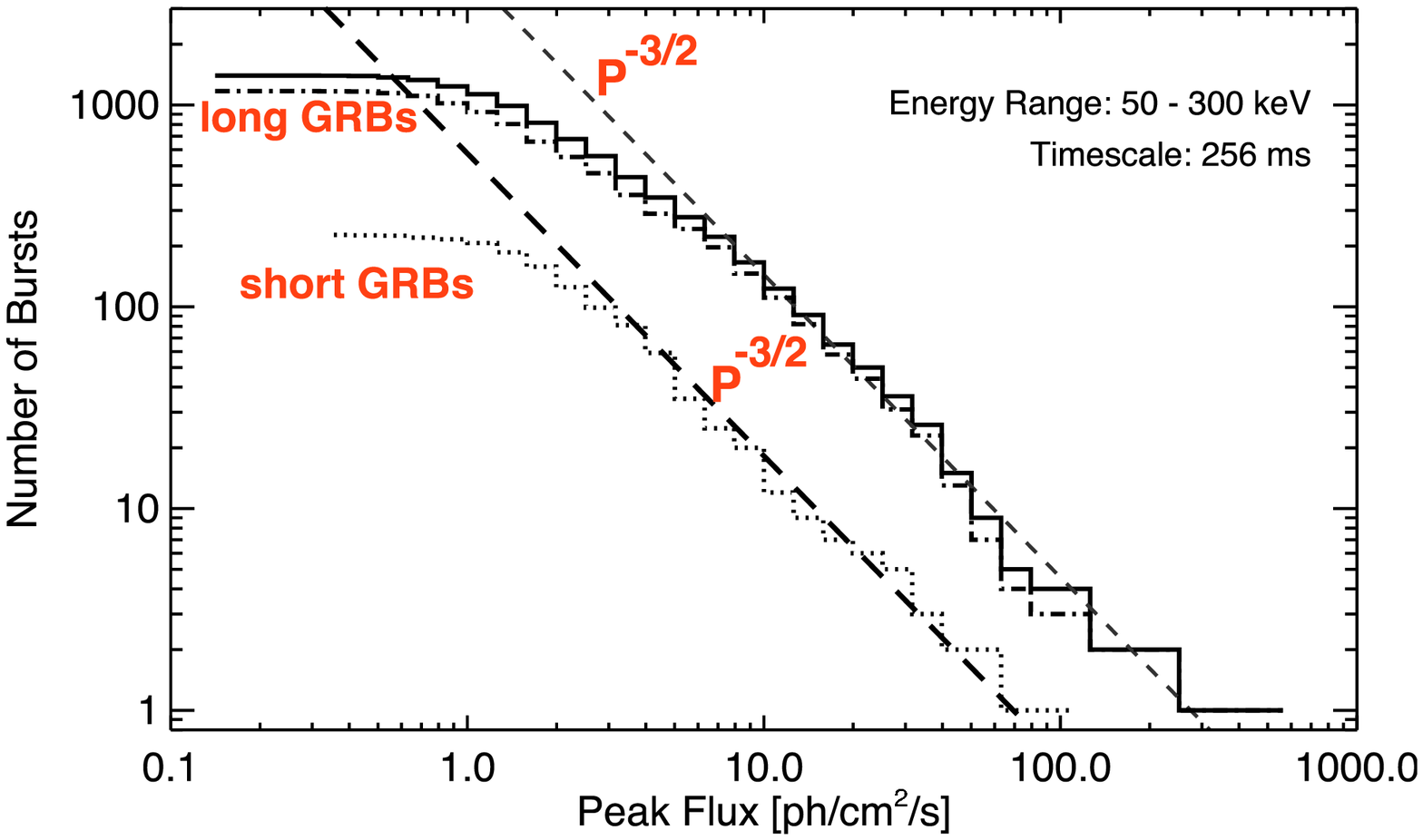}
\caption{\label{pf256_fig} Same as Figure~\ref{pflx_fig}, except on the 0.256~s timescale.}
\end{center}
\end{figure}

\begin{figure}
\begin{center}
 \epsscale{0.95}
\plotone{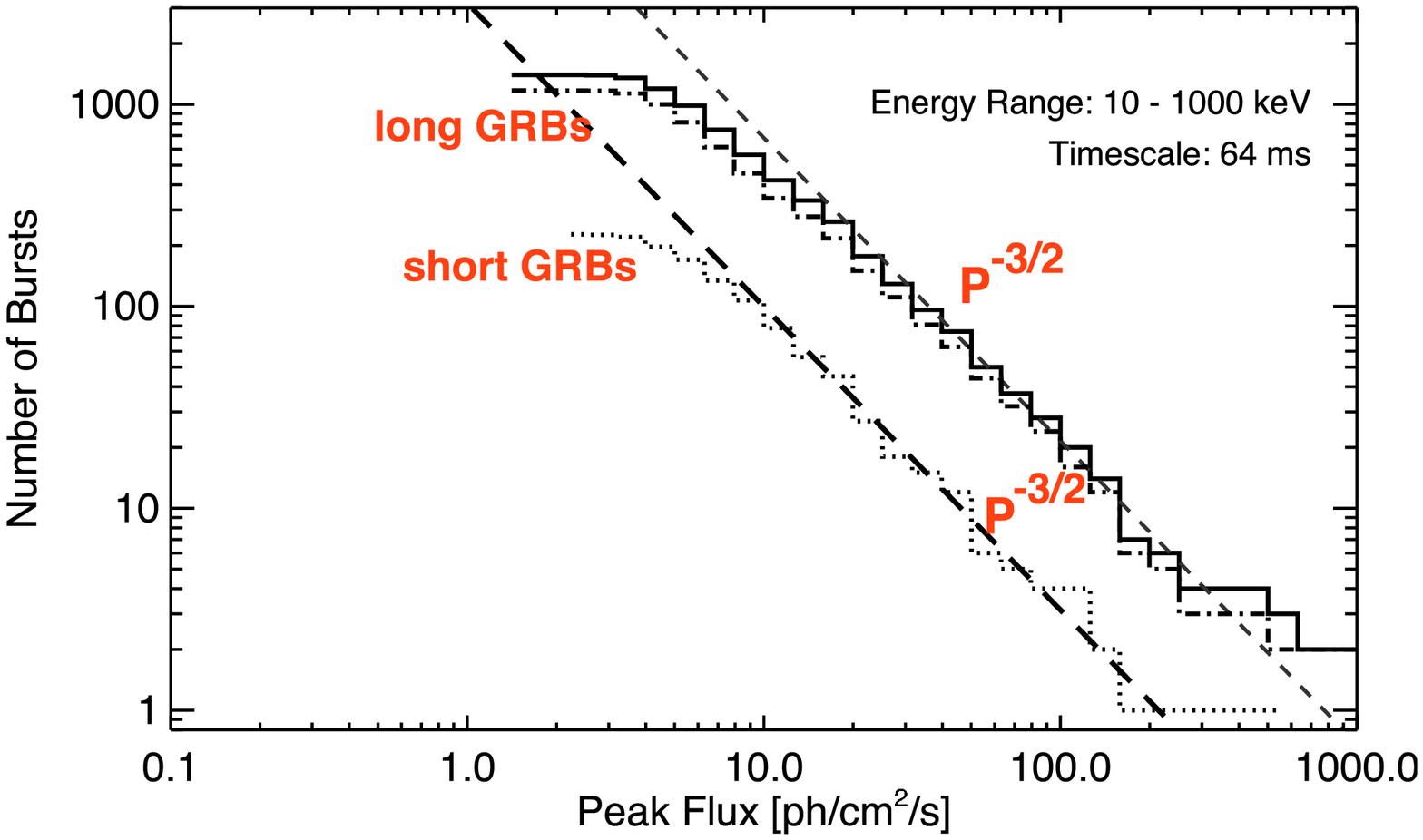}
\plotone{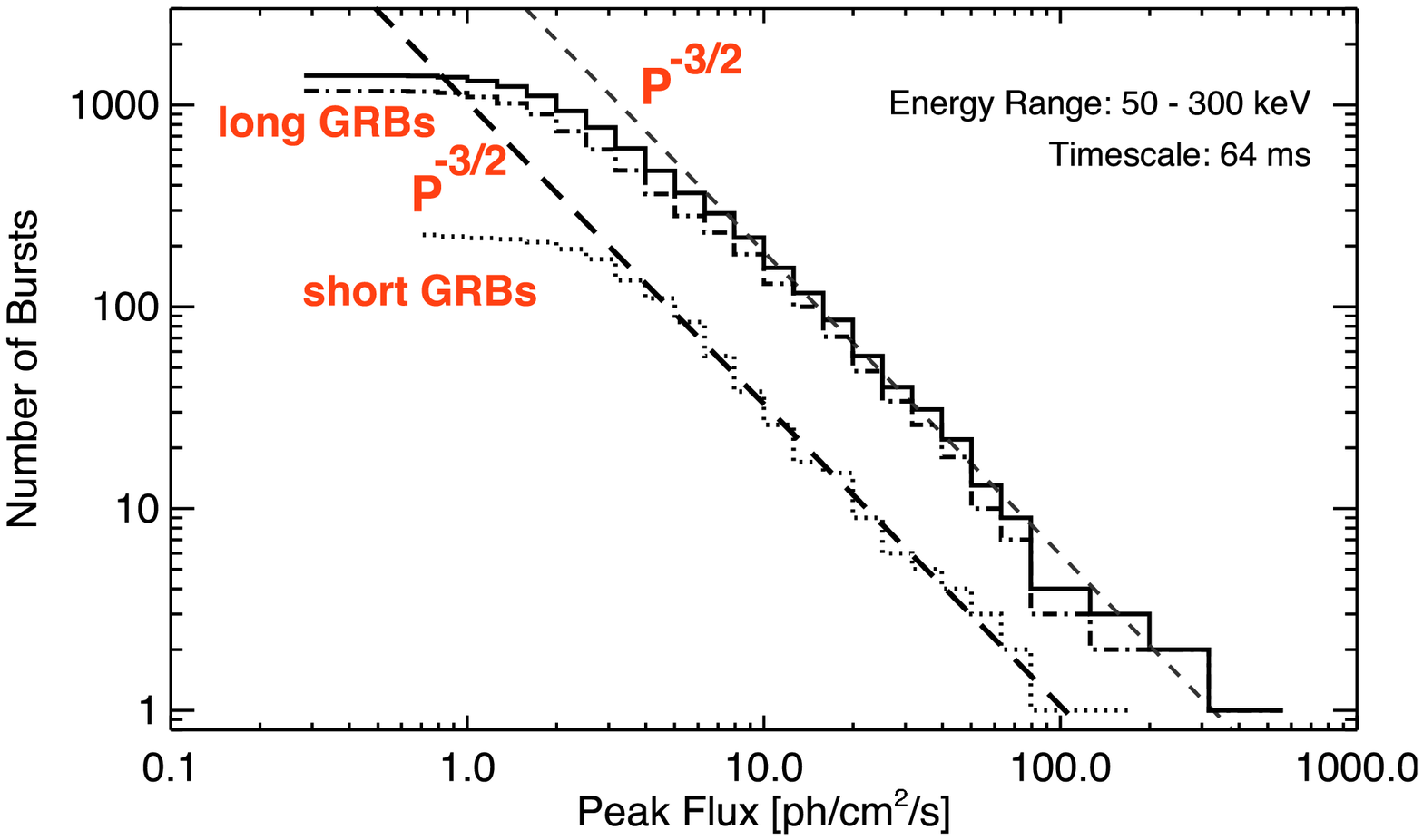}
\caption{\label{pf64_fig}Same as Figure~\ref{pflx_fig}, except on the 0.064~s timescale.}
\end{center}
\end{figure}

\begin{figure}
\begin{center}
 \epsscale{0.95}
\plotone{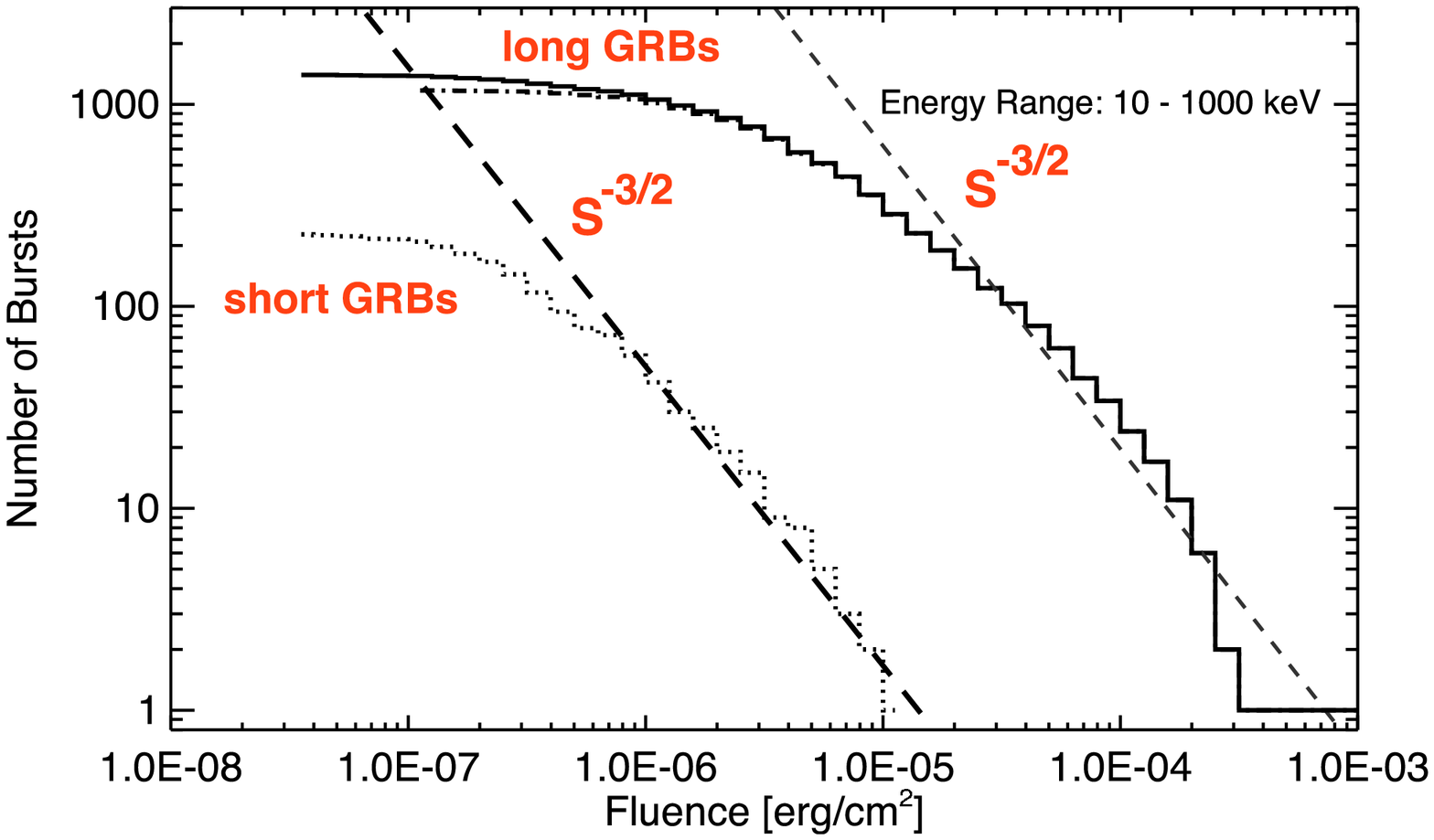}
\plotone{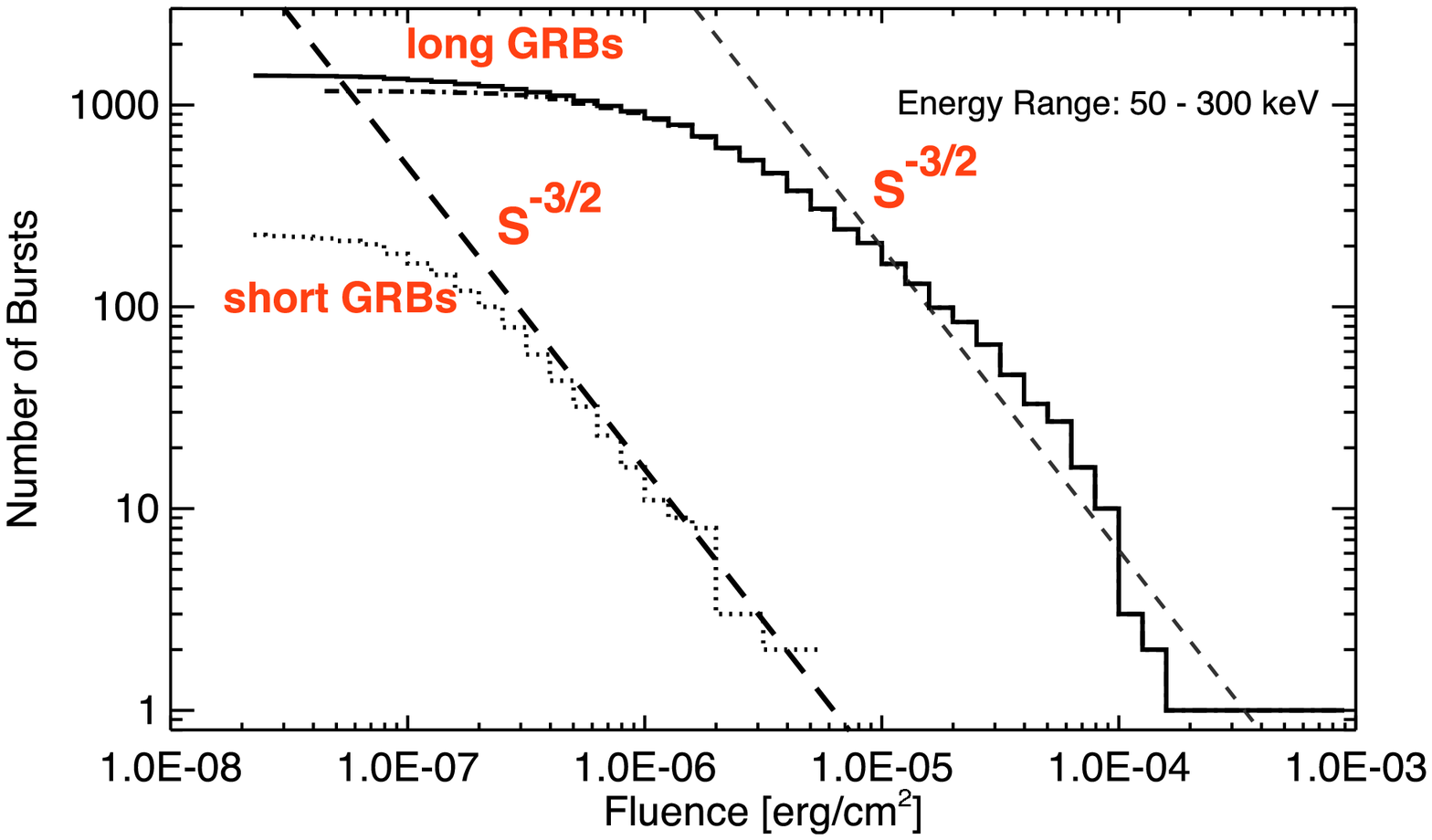}
\caption{\label{flu_fig} Integral distribution of GRB fluence (S ergs/cm$^2$) in two energy ranges: 10--1000~keV (upper plot) and 50--300 keV (lower plot). Distributions are shown for the total sample (solid histogram), short GRBs (dots) and long GRBs (dash-dots), using $T_{90} = 2$~s as the distinguishing criterion. In each plot a power law with a slope of $-3 / 2$ (dashed line) is drawn to guide the eye.}
\end{center}
\end{figure}

Integral distributions of the peak fluxes observed for GRBs in the first six years are shown in Figures~\ref{pflx_fig} -- \ref{pf64_fig} for the three different timescales and separately for short and long GRBs. 
In a Euclidean space, if all the long GRB progenitors are uniformly distributed, they would be expected to follow a -3/2 power-law. However, in a Big Bang cosmology, the effects of the (unknown) GRB luminosity function, along with their distribution in redshift (which follows more or less the star formation rate up to z$\sim$4) and the threshold of the instrument are all convolved together to produce the observed distribution.
The GRB peak flux distributions could however provide useful constraints in various astrophysical studies, such as in determining the true GRB rate for any jet model \citep{2005ApJ...622..482G}. The integral fluence distributions for the two energy intervals are also shown in Figure~\ref{flu_fig}. It may be noted that the integral fluence distributions show far more curvature than the integral peak flux distributions. Although the deficit at the highest fluence end may be due to small number statistics, the observed departure from the expected power-law of slope -3/2 at low fluence may be due to the fact that these GRB fluence estimates suffer from relatively large systematic errors, primarily due to limited spectral channels and rather narrow time bins with very few events that are used for spectral fitting. Moreover, GRB triggers on GRB peak flux and not on fluence, so that there is no clear well defined GBM fluence threshold. 

\section{SUMMARY}
\label{CatSum}

The third GBM catalog comprises a list of 1403 cosmic GRBs that triggered GBM between July 12, 2008 and  July 11, 2014. The increased sample of GRB  in this catalog confirms the conclusions of the earlier two year and four year catalogs.
The 6 year average GRB rate detected by GBM is (0.662 $\pm$ 0.018) per day. The shortest and longest time interval between consecutive GRB triggers is 10 minutes and 11.77 days (with no interruptions) respectively. The shortest time interval is close to the minimum time for GBM to re-trigger. The longest time interval is $\sim$\ 8 times larger than the average value of 1.5 days. Assuming that the GRB triggers follow Poisson statistics, this implies that such a large time gap could occur once in $\sim$ 10 years.
The average rate of burst detections per year ($\sim 242 \pm 6.5$/year), which is only slightly smaller compared to the detection rate of the BATSE instrument of $\sim 300$/year \citep{1999ApJS..122..465P}. One would have expected a larger difference due to the superior detection sensitivity of BATSE. This perhaps can be explained by GBM's additional range of trigger timescales (primarily the 2~s and 4~s timescales), which are compensating for the higher burst detection threshold of GBM  ($\sim 0.7$ vs. $\sim 0.2$ photons cm$^{-2}$ s$^{-1}$ for BATSE). The distribution of GBM durations is consistent with the well-known bimodal distribution measured previously \citep{CK93}. This is supported by two independent analyses. Assuming a $T_{90}$ value of 2 s to distinguish between short and long GRBs, the median durations of the two groups are 0.58\ s and 26.62\ s for the GBM sample while they are 0.45\ s and 30.75\ s  respectively for the BATSE sample.  The median duration of GBM detected long bursts is lower than that of BATSE detected long GRBs. However the fraction of short GRBs in the GBM sample is about 21\% which is not significantly different from that detected by BATSE which is 24\%. 

\acknowledgments

Support for the German contribution to GBM was provided by the Bundesministerium f\"ur Bildung und Forschung (BMBF) via the Deutsches Zentrum f\"ur Luft und Raumfahrt (DLR) under contract number 50 QV 0301. A.v.K. was supported by the Bundesministeriums f\"ur Wirtschaft und Technologie (BMWi) through DLR grant 50 OG 1101. SMB and OJR acknowledge support from Science Foundation Ireland under Grant No. 12/IP/1288.
HFY acknowledges support by the DFG cluster of excellence ``Origin and Structure of the Universe". A.G. is funded by the NASA Postdoctoral Program through Oak Ridge Associated Universities. The UAH co-authors gratefully acknowledge NASA funding from co-operative agreement NNM11AA01A. CK and CAWH gratefully acknowledge NASA funding through the \Fermi\ GBM project.

\clearpage

 \clearpage

\clearpage

\clearpage

 \clearpage

 \clearpage
 
 \clearpage

 \clearpage
 
\clearpage

 \clearpage

\clearpage

\clearpage

\clearpage

 \clearpage

\clearpage

\clearpage

\begin{table}[h]
  \centering
  \caption{Trigger statistics of the year 1 \& 2 and year 3 \& 4 catalogs }\label{trigstat1st2ndcat}
\begin{tabular}{|l||c|c|c|c|c|c|c|c|c|}
  \hline
   & GRBs & SGRs & TGFs & SFs & CPs & Other & Sum & ARRs & LAT GRBs \\ \hline
  Year 1 \& 2 & 492\tablenotemark{a} & 171 & 79 & 31 & 68\tablenotemark{b} & 65\tablenotemark{b} & 906\tablenotemark{c} & 40 & 22 \\ \hline
  Year 3 \& 4  & 462 & 18 & 183 & 363 & 141 & 53 & 1220 & 46 & $21\tablenotemark{d}$  \\ \hline
  Year 5 \& 6  & 451\tablenotemark{a} & 9 & 207 &400 & 96  & 61& 1224 &33  & $ 29 \tablenotemark{d}$   \\ \hline
  Year 1 to 6  & 1405&198 & 469 & 794 & 305 & 179 & 3350 & 119\tablenotemark{e} & 72 \\
  \hline
\end{tabular}
\tablenotetext{a}{GRB 091024A and GRB130925A each of which triggered GBM twice, are counted twice. Hence the total number of GRB's is one less in each group.}
\tablenotetext{b}{ The numbers of non GRB triggers in year 1 \& 2 differ from the numbers cited in \cite{Pacie12}, since some of the triggers were reclassified}
\tablenotetext{c}{The total numbers of triggers is two less compared to \cite{Pacie12}, since the two commanded triggers (bn100709294 \& bn100711145) were not counted.}
\tablenotetext{d}{The three year \Fermi\ LAT GRB catalog \citep{2013ApJS..209...11A} includes bursts only from August 2008 to August 2011 (Year 1 \& 2: 22 GRBs, Year 3 \& 4: 13 GRBs). The 29 additional GRB detections from year 5 \& 6, are listed in the public GRB list of the \Fermi\ LAT team: \url{http://fermi.gsfc.nasa.gov/ssc/observations/types/grbs/lat_grbs/}}
\tablenotetext{e}{Due to misclassification of events as GRBs by the FSW, 26 of the ARRs occurred for other event types. Of these 16 occurred due to charged particle events, 5 occurred due to SGR events, 4 occurred due to solar flare events and 1  due to a TGF event. In addition, there were a few positive ARRs from GBM triggers followed by no spacecraft slews, which was disabled at the spacecraft level at that time. In a few cases the spacecraft slew started well after the GBM trigger due to Earth's limb constraint.}
\end{table}




\begin{deluxetable}{ccccccccccccc}

\rotate

\tabletypesize{\small}

\tablewidth{614pt}

\tablecaption{\label{trigger:criteria:history} Trigger Criteria History}


\tablehead{\colhead{Algorithm} & \colhead{Timescale} & \colhead{Offset} & \colhead{Channels} & \colhead{Energy} & \multicolumn{7}{c}{Threshold ($0.1 \sigma$)\tablenotemark{a}}\\
\cline{6-13}\\
\multicolumn{5}{c}{ } & \multicolumn{3}{c}{ 2008} & \multicolumn{4}{c}{2009 } & 2010\\
\colhead{Number} & \colhead{(ms)} & \colhead{(ms)} & \colhead{} & \colhead{(keV)} & \colhead{July 11} & \colhead{July 14} & \colhead{Aug 1} & \colhead{May 8} & \colhead{Oct 29 } & \colhead{Nov 10} & \colhead{Dec 7} & \colhead{Mar 26}}
\startdata
1 & 16 & 0 & 3--4 & 50--300 & 75 & 24 & 24 & 24 & 24 & 24 & 24 & 24 \\
2 & 32 & 0 & 3--4 & 50--300 & 75 & 24 & 24 & 24 & 24 & 24 & 24 & 24 \\
3 & 32 & 16 & 3--4 & 50--300 & 75 & 24 & 24 & 24 & 24 & 24 & 24 & 24 \\
4 & 64 & 0 & 3--4 & 50--300 & 45 & 24 & 50 & 24 & 24 & 24 & 24 & 24 \\
5 & 64 & 32 & 3--4 & 50--300 & 45 & 24 & 50 & 24 & 24 & 24 & 24 & 24 \\
6 & 128 & 0 & 3--4 & 50--300 & 45 & 24 & 48 & 50 & 24 & 24 & 24 & 24 \\
7 & 128 & 64 & 3--4 & 50--300 & 45 & 24 & 48 & 50 & 24 & 24 & 24 & 24 \\
8 & 256 & 0 & 3--4 & 50--300 & 45 & 24 & 24 & 24 & 24 & 24 & 24 & 24 \\
9 & 256 & 128 & 3--4 & 50--300 & 45 & 24 & 24 & 24 & 24 & 24 & 24 & 24 \\
10 & 512 & 0 & 3--4 & 50--300 & 45 & 24 & 24 & 24 & 24 & 24 & 24 & 24 \\
11 & 512 & 256 & 3--4 & 50--300 & 45 & 24 & 24 & 24 & 24 & 24 & 24 & 24 \\
12 & 1024 & 0 & 3--4 & 50--300 & 45 & 24 & 24 & 24 & 24 & 24 & 24 & 24 \\
13 & 1024 & 512 & 3--4 & 50--300 & 45 & 24 & 24 & 24 & 24 & 24 & 24 & 24 \\
14 & 2048 & 0 & 3--4 & 50--300 & 45 & 24 & 24 & 24 & 24 & 24 & 24 & 24 \\
15 & 2048 & 1024 & 3--4 & 50--300 & 45 & 24 & 24 & 24 & 24 & 24 & 24 & 24 \\
16 & 4096 & 0 & 3--4 & 50--300 & 45 & 24 & 24 & 24 & 24 & 24 & 24 & 24 \\
17 & 4096 & 2048 & 3--4 & 50--300 & 45 & 24 & 24 & 24 & 24 & 24 & 24 & 24 \\
18 & 8192 & 0 & 3--4 & 50--300 & C & 50 & 24 & 24 & D & 24 & 24 & 24 \\
19 & 8192 & 4096 & 3--4 & 50--300 & C & 50 & 24 & 24 & D & 24 & 24 & 24 \\
20 & 16384 & 0 & 3--4 & 50--300 & C & 50 & D & 24 & 24 & 24 & 24 & 24 \\
21 & 16384 & 8192 & 3--4 & 50--300 & C & 50 & D & 24 & 24 & 24 & 24 & 24 \\
22 & 16 & 0 & 2--2 & 25--50 & D & 80 & 24 & 24 & 24 & 24 & 24 & 24 \\
23 & 32 & 0 & 2--2 & 25--50 & D & 80 & 24 & 24 & 24 & 24 & 24 & 24 \\
24 & 32 & 16 & 2--2 & 25--50 & D & 80 & 24 & 24 & 24 & 24 & 24 & 24 \\
25 & 64 & 0 & 2--2 & 25--50 & D & 55 & 24 & 24 & 24 & 24 & 24 & 24 \\
26 & 64 & 32 & 2--2 & 25--50 & D & 55 & 24 & 24 & 24 & 24 & 24 & 24 \\
27 & 128 & 0 & 2--2 & 25--50 & D & 55 & 24 & 24 & D & 24 & 24 & 24 \\
28 & 128 & 64 & 2--2 & 25--50 & D & 55 & 24 & 24 & D & 24 & 24 & 24 \\
29 & 256 & 0 & 2--2 & 25--50 & D & 55 & 24 & 24 & D & 24 & 24 & 24 \\
30 & 256 & 128 & 2--2 & 25--50 & D & 55 & 24 & 24 & D & 24 & 24 & 24 \\
31 & 512 & 0 & 2--2 & 25--50 & D & 55 & 24 & 24 & D & 24 & 24 & 24 \\
32 & 512 & 256 & 2--2 & 25--50 & D & 55 & 24 & 24 & D & 24 & 24 & 24 \\
33 & 1024 & 0 & 2--2 & 25--50 & D & 55 & 24 & 24 & D & 24 & 24 & 24 \\
34 & 1024 & 512 & 2--2 & 25--50 & D & 55 & 24 & 24 & D & 24 & 24 & 24 \\
35 & 2048 & 0 & 2--2 & 25--50 & D & 55 & 24 & 24 & D & 24 & 24 & 24 \\
36 & 2048 & 1024 & 2--2 & 25--50 & D & 55 & 24 & 24 & D & 24 & 24 & 24 \\
37 & 4096 & 0 & 2--2 & 25--50 & D & 65 & 24 & 24 & D & 24 & 24 & 24 \\
38 & 4096 & 2048 & 2--2 & 25--50 & D & 65 & 24 & 24 & D & 24 & 24 & 24 \\
39 & 8192 & 0 & 2--2 & 25--50 & D & 65 & 24 & 24 & D & 24 & 24& 24  \\
40 & 8192 & 4096 & 2--2 & 25--50 & D & 65 & 24 & 24 & D & 24 & 24 & 24 \\
41 & 16384 & 0 & 2--2 & 25--50 & D & 65 & D & 24 & 24 & 24 & 24 & 24 \\
42 & 16384 & 8192 & 2--2 & 25--50 & D & 65 & D & 24 & 24 & 24 & 24 & 24 \\
43 & 16 & 0 & 5--7 & $> 300$ & D & 80 & 24 & 24 & 24 & 24 & 24 & 24 \\
44 & 32 & 0 & 5--7 & $> 300$ & D & 80 & 24 & 24 & D & 24 & 24 & 24 \\
45 & 32 & 16 & 5--7 & $> 300$ & D & 80 & 24 & 24 & D & 24 & 24 & 24 \\
46 & 64 & 0 & 5--7 & $> 300$ & D & 55 & 24 & 60 & D & 24 & 24 & 24 \\
47 & 64 & 32 & 5--7 & $> 300$ & D & 55 & 24 & 60 & D & 24 & 24 & 24 \\
48 & 128 & 0 & 5--7 & $> 300$ & D & 55 & 24 & 24 & D & 24 & 24 & 24 \\
49 & 128 & 64 & 5--7 & $> 300$ & D & 55 & 24 & 24 & D & 24 & 24 & 24 \\
50 & 16 & 0 & 4--7 & $> 100$ & D & 80 & 24 & 24 & 24 & 24 & 24 & 24 \\
51 & 32 & 0 & 4--7 & $> 100$ & D & 80 & 24 & 24 & D & 24 & 24 & 24 \\
52 & 32 & 16 & 4--7 & $> 100$ & D & 80 & 24 & 24 & D & 24 & 24 & 24 \\
53 & 64 & 0 & 4--7 & $> 100$ & D & 55 & 24 & 24 & D & 24 & 24 & 24 \\
54 & 64 & 32 & 4--7 & $> 100$ & D & 55 & 24 & 24 & D & 24 & 24 & 24 \\
55 & 128 & 0 & 4--7 & $> 100$ & D & 55 & 24 & 24 & D & 24 & 24 & 24 \\
56 & 128 & 64 & 4--7 & $> 100$ & D & 55 & 24 & 24 & D & 24 & 24 & 24 \\
57 & 256 & 0 & 4--7 & $> 100$ & D & 55 & 24 & 24 & D & 24 & 24 & 24 \\
58 & 256 & 128 & 4--7 & $> 100$ & D & 55 & 24 & 24 & D & 24 & 24 & 24 \\
59 & 512 & 0 & 4--7 & $> 100$ & D & 55 & 24 & 24 & D & 24 & 24 & 24 \\
60 & 512 & 256 & 4--7 & $> 100$ & D & 55 & 24 & 24 & D& 24  & 24 & 24 \\
61 & 1024 & 0 & 4--7 & $> 100$ & D & 55 & 24 & 24 & D & 24 & 24 & 24 \\
62 & 1024 & 512 & 4--7 & $> 100$ & D & 55 & 24 & 24 & D & 24 & 24 & 24 \\
63 & 2048 & 0 & 4--7 & $> 100$ & D & 55 & 24 & 24 & D & 24 & 24 & 24 \\
64 & 2048 & 1024 & 4--7 & $> 100$ & D & 55 & 24 & 24 & D & 24 & 24 & 24 \\
65 & 4096 & 0 & 4--7 & $> 100$ & D & 65 & 24 & 24 & D & 24 & 24 & 24 \\
66 & 4096 & 2048 & 4--7 & $> 100$ & D & 65 & 24 & 24 & D & 24 & 24 & 24 \\
&   &  & 5--7 & $> 300$  &  &  &  &  &  & 60 & 55 & : \\
\rb{116\tablenotemark{b}} & \rb{16} & \rb{0} & BGO/3--6 & 2 - 40 MeV & \rb{D} & \rb{:} & \rb{:} & \rb{:} & \rb{:} & 55 & 45 & : \\
 &   &  & 5--7 & $> 300$ &  &  &  &  & & 55 & 45 & : \\
\rb{117\tablenotemark{b}} & \rb{16} & \rb{0} & BGO/3--6 & 2 - 40 MeV & \rb{D} & \rb{:} & \rb{:} & \rb{:} & \rb{:} & 55 & 45 & : \\
 &   &  & 5--7 & $> 300$ &  &  &  &  & & 55 & 45 & : \\
\rb{118\tablenotemark{b}} & \rb{16} & \rb{0} & BGO/3--6 &  2 - 40 MeV & \rb{D} & \rb{:} & \rb{:} & \rb{:} & \rb{:} & 55 & 45 & : \\
\raisebox{0.0ex}{119\tablenotemark{b}} & 16 & 0 & BGO/3--6 &  2 - 40 MeV  & D & : & : & : & : & 55 & 45 & 47\\
\enddata
\tablenotetext{a}{Symbol ':' indicates no change from previous setting; 'C' indicates that the algorithm is in compute mode (see text); 'D' indicates that the algorithm is disabled.}
\tablenotetext{b}{Trigger algorithms using the BGO detector count rates. Algorithm 116 triggers off when at least two NaI and  one BGO detectors are exceeding the trigger threshold.  Algorithms 117 is same as 116, but impose the additional requirement that the triggered detectors are on the +X side of the spacecraft.  Algorithm 118 is the same as 117, but requiring the triggered detectors to be on the -X side of the spacecraft. Algorithm 119 requires a significant rate increase in both BGO detectors.}


\end{deluxetable}

\begin{table}[h]
  \centering
  \caption{Trigger algorithm statistics}\label{trigstatalgor}
\begin{tabular}{|c|c|c|c|c|c|c|c|c|c|}
  \hline
  Algorithm & Time ms & Energy keV & GRBs & SGRs & TGFs & SFs & CPs & Other & Comment\tablenotemark{a} \\ \hline
 1 - 5 & 16 -64  & 50 - 300        & 235 & 73 & 8   &  6  & 9   & 17 & GRB \\ \hline
 6 - 11 & 128 - 512  &  50 - 300  & 509 & 8  & -   & 41  & 12  & 37 & GRB \\ \hline
  12 - 17 & 1024 - 4096  &  50 - 300  & 639 & 1  & -   & 167  & 230  & 30 & GRB\\ \hline
  18 - 21 & 8192 - 16384 &  50 - 300 & 3 & -  & -   & -  & - & - & D\\ \hline
  22 - 26 & 16 -64 & 25 - 50        & 7   & 113 & -   & 577 & 11   &  7 & SF\\ \hline
  27 - 32 & 128 - 512  & 25 - 50   & 2  & 3  & -   &  1  & 2   &  - & D\\ \hline
  33 - 38 & 1024 - 4096 & 25 - 50   & 8   & -  & -   &  2  & 11   &  3 & D\\ \hline
  39 - 42 &  8192 - 16384 & 25 - 50   & 1   & -  & -   &  -  & 8   &  3 & D\\ \hline
    43  & 16 &$> 300$         & -   & -  & 43  &  -  & 1   &  1 & TGF\\ \hline
  44 - 49 & 32 - 128&$> 300$         & -   & -  & -  &  -  & -   &  5 & D\\ \hline
  50  & 16  & $> 100$      & -   & -  & 9   &  -  & 4   &  - & TGF\\ \hline
  51 - 66 & 32 - 4096 & $> 100$      & 1   & -  &  -  &  -  &  -   &  - & D\\ \hline
 116 - 119 & 16 & BGO               & -   & -  & 409 &  -  & 16   &  76 & TGF\\ \hline
\end{tabular}

\tablenotetext{a}{'GRB', 'SF' and 'TGF' indicate the source classes that are most likely to trigger the corresponding algorithm; 'D' indicates that the algorithm was finally disabled at the end of year 4.}
\end{table}




 commands


\end{deluxetable}




 commands
\tablenotetext{a}{Other instrument detections: Mo: Mars Observer , K: Konus-Wind, R: RHESSI, IA: \INTEGRAL\ SPI-ACS, IS: \INTEGRAL\ IBIS-ISGRI, S: \Swift, Me: Messenger, W: \Suzaku, A: \AGILE, M: \MAXI, L: \Fermi\ LAT, Nu: NuSTAR, iPTF: intermediate Palomar Transient Factory, ARR: Autonomous Repoint Requests by GBM FSW}
\tablenotetext{aa}{GRB120801 There is a delayed emission at T0+$\sim$400s.}
\tablenotetext{b}{GRB091024A triggered GBM twice.}
\tablenotetext{bb}{GRB121123A GBM did not trigger on pre-trigger which triggered Swift; T90 is incorrect}
\tablenotetext{c}{GRB130307A possible precursors of this trigger were unobservable since it triggered soon after SAA exit.}
\tablenotetext{cc}{GRB121217A Swift triggered $\sim$12 min before T0. This GRB has several episodes well separated in time. Hence T$_{90}$ is possibly incorrect}
\tablenotetext{d}{GRB130604B Fermi enters SAA $\sim$ 105s after trigger.}
\tablenotetext{dd}{GRB131028 This GRB triggrered during a X1.0 Solar Flare.}
\tablenotetext{e}{GRB130907 Fermi enters SAA $\sim$ 130s after trigger.}
\tablenotetext{ee}{GRB131108 A second GRB131108A occurred $\sim$ 225s after this GRB triggered.}
\tablenotetext{f}{GRB130909 Fermi enters SAA $\sim$ 53s after trigger.}
\tablenotetext{ff}{GRB131123 This GRB triggrered during a M1.0 Solar Flare.}
\tablenotetext{g}{GRB130925A triggered GBM twice.}
\tablenotetext{h}{GRB140115 Fermi enters SAA $\sim$ 50s after trigger.}
\tablenotetext{i}{GRB130206A Swift-BAT triggered at 07:17:20 UT on first emission period of GRB1140206A, GBM on the second pulse  at  Swift T0+56s.}
\tablenotetext{j}{GRB140219 Fermi enters SAA $\sim$ 9s after trigger.}
\tablenotetext{k}{GRB140329A Fermi enters SAA $\sim$ 120s after trigger.}
\tablenotetext{l}{GRB140404 There is a precursor at T0-~70s.}
\tablenotetext{m}{GRB140329 Fermi enters SAA $\sim$ 155s after trigger.}
\tablenotetext{n}{GRB140329A Fermi enters SAA $\sim$ 60s after trigger.}
\tablenotetext{o}{GRB140517 Fermi enters SAA $\sim$ 65s after trigger.}
\tablenotetext{p}{GRB140627 Fermi enters SAA $\sim$ 190s after trigger.}


\end{deluxetable}




 commands
\tablenotetext{a}{Data problems precluded duration analysis.}
\tablenotetext{b}{Used TTE binned at 32 ms.}
\tablenotetext{c}{Partial earth occultation is likely; durations are lower limits.}
\tablenotetext{d}{Possible precursor at $\sim T_0-120$ s.}
\tablenotetext{e}{Data cut off due to SAA entry while burst in progress; durations are lower limits.}
\tablenotetext{f}{SAA entry at $T_0+83$~s; durations are lower limits.}
\tablenotetext{g}{Used TTE binned at 16 ms.}
\tablenotetext{h}{This GRB triggered GBM twice.}
\tablenotetext{i}{Too weak to measure durations; visual duration is $\sim 0.025$~s.}
\tablenotetext{j}{Possible contamination due to emergence of Crab \& A0535+26 from Earth occultation.}
\tablenotetext{k}{Solar activity starting at $T_0+200$~s. Post burst background interval was selected  before.}
\tablenotetext{l}{Data cut off due to SAA entry while burst in progress;it is not possible to determine durations.}
\tablenotetext{m}{Spacecraft in sun pointing mode, detector threshold raised, location of burst nearly in -z direction. The response, peak fluxes and fluence in the $10 -100$~keV energy range have large errors. Fluence, peak fluxes and durations in BATSE energy range (50 -300 keV) are reliable.}
\tablenotetext{n}{{L}ocalization of precursor at $T_0-120$~s is consistent with burst location and was included in the duration analysis.}
\tablenotetext{o}{SAA entry at $T_0+100$~s; durations are lower limits.}
\tablenotetext{p}{TTE/CTTE data not available, 64~ms peak fluxes may not be correct.}



\end{deluxetable}




 commands



\end{deluxetable}

\end{document}